%% file: main.tex
\documentclass[sigconf,nonacm]{acmart}

\AtBeginDocument{%
  }

\copyrightyear{2025}
\acmYear{2025}
\setcopyright{cc}
\setcctype{by}
\acmConference[FAccT '25]{The 2025 ACM Conference on Fairness, Accountability, and Transparency}{June 23--26, 2025}{Athens, Greece}
\acmBooktitle{The 2025 ACM Conference on Fairness, Accountability, and Transparency (FAccT '25), June 23--26, 2025, Athens, Greece}\acmDOI{10.1145/3715275.3732063}
\acmISBN{979-8-4007-1482-5/2025/06}

\input{utils/utils}

\begin{document}

\title{From Lived Experience to Insight: Unpacking the Psychological Risks of Using AI Conversational Agents}

\author{Mohit Chandra}
\authornote{Work done during internship at Microsoft Research.}
\authornote{Corresponding Author}
\email{mchandra9@gatech.edu}
\orcid{0000-0001-5030-6647}
\affiliation{%
  \institution{School of Interactive Computing, Georgia Institute of Technology}
  \city{Atlanta}
  \state{Georgia}
  \country{USA}
}

\author{Suchismita Naik}
\authornotemark[1]
\email{naik33@purdue.edu}
\orcid{0009-0002-5667-4576}
\affiliation{%
  \institution{Purdue University}
  \city{West Lafayette}
  \state{Indiana}
  \country{USA}
}

\author{Denae Ford}
\email{denae@microsoft.com}
\orcid{0000-0003-0654-4335}
\affiliation{%
  \institution{Microsoft Research}
  \city{Redmond}
  \state{Washington}
  \country{USA}
}

\author{Ebele Okoli}
\email{ebeleokoli@microsoft.com}
\orcid{0009-0002-3689-0490}
\affiliation{%
  \institution{Microsoft}
  \city{Atlanta}
  \state{Georgia}
  \country{USA}
}

\author{Munmun De Choudhury}
\email{munmun.choudhury@cc.gatech.edu}
\orcid{0000-0002-8939-264X}
\affiliation{%
   \institution{School of Interactive Computing, Georgia Institute of Technology}
  \city{Atlanta}
  \state{Georgia}
  \country{USA}
}

\author{Mahsa Ershadi}
\email{mahsaershadi@microsoft.com}
\orcid{0000-0002-7660-6442}
\affiliation{%
  \institution{Microsoft AI}
  \city{Vancouver}
  \state{British Columbia}
  \country{Canada}
}

\author{Gonzalo Ramos}
\email{goramos@microsoft.com}
\orcid{0000-0003-4198-5021}
\affiliation{%
  \institution{Microsoft Research}
  \city{Redmond}
  \state{Washington}
  \country{USA}
}

\author{Javier Hernandez}
\email{javierh@microsoft.com}
\orcid{0000-0001-9504-5217}
\affiliation{%
  \institution{Microsoft Research}
  \city{Redmond}
  \state{Washington}
  \country{USA}
}

\author{Ananya Bhattacharjee}
\authornotemark[1]
\email{ananya@cs.toronto.edu}
\orcid{0000-0002-9116-3766}
\affiliation{%
  \institution{University of Toronto}
  \city{Toronto}
  \state{Ontario}
  \country{Canada}
}

\author{Shahed Warreth}
\email{swarreth@microsoft.com}
\orcid{0000-0001-5843-1565}
\affiliation{%
  \institution{Microsoft AI}
  \city{Dublin}
  \country{Ireland}
}

\author{Jina Suh}
\authornotemark[2]
\email{jinsuh@microsoft.com}
\orcid{0000-0002-7646-5563}
\affiliation{%
  \institution{Microsoft Research}
  \city{Redmond}
  \state{Washington}
  \country{USA}
}

\renewcommand{\shortauthors}{Chandra, et al.}

\begin{abstract}
  \input{00_abstract}
\end{abstract}



\maketitle

\input{01_introduction}
\input{02_related_work}

\input{03_phase1}
\input{04_phase2}

\input{05_discussion}

\bibliographystyle{BibFiles/ACM-Reference-Format}
\bibliography{BibFiles/references}

\appendix
\input{06_appendix}

\end{document}

%% file: utils/utils.tex
\newcommand{\subcategory}[1]{\textit{#1}}

%% file: 00_abstract.tex
Recent gains in popularity of AI conversational agents have led to their increased use for improving productivity and supporting well-being. While previous research has aimed to understand the risks associated with interactions with AI conversational agents, these studies often fall short in capturing the lived experiences of individuals. Additionally, psychological risks have often been presented as a sub-category within broader AI-related risks in past taxonomy works, leading to under-representation of the impact of psychological risks of AI use. To address these challenges, our work presents a novel risk taxonomy focusing on psychological risks of using AI gathered through the lived experiences of individuals. We employed a mixed-method approach, involving a comprehensive survey with 283 people with lived mental health experience and workshops involving experts with lived experience to develop a psychological risk taxonomy. Our taxonomy features 19 AI behaviors, 21 negative psychological impacts, and 15 contexts related to individuals. Additionally, we propose a novel multi-path vignette-based framework for understanding the complex interplay between AI behaviors, psychological impacts, and individual user contexts. Finally, based on the feedback obtained from the workshop sessions, we present design recommendations for developing safer and more robust AI agents. Our work offers an in-depth understanding of the psychological risks associated with AI conversational agents and provides actionable recommendations for policymakers, researchers, and developers.

\noindent\textcolor{red}{\textbf{Content Warning:} This paper includes discussions of sensitive topics, including but not limited to self-harm, body shaming, and discrimination.}

%% file: 01_introduction.tex
\section{Introduction}

\begin{figure*}
  \includegraphics[width=\textwidth]{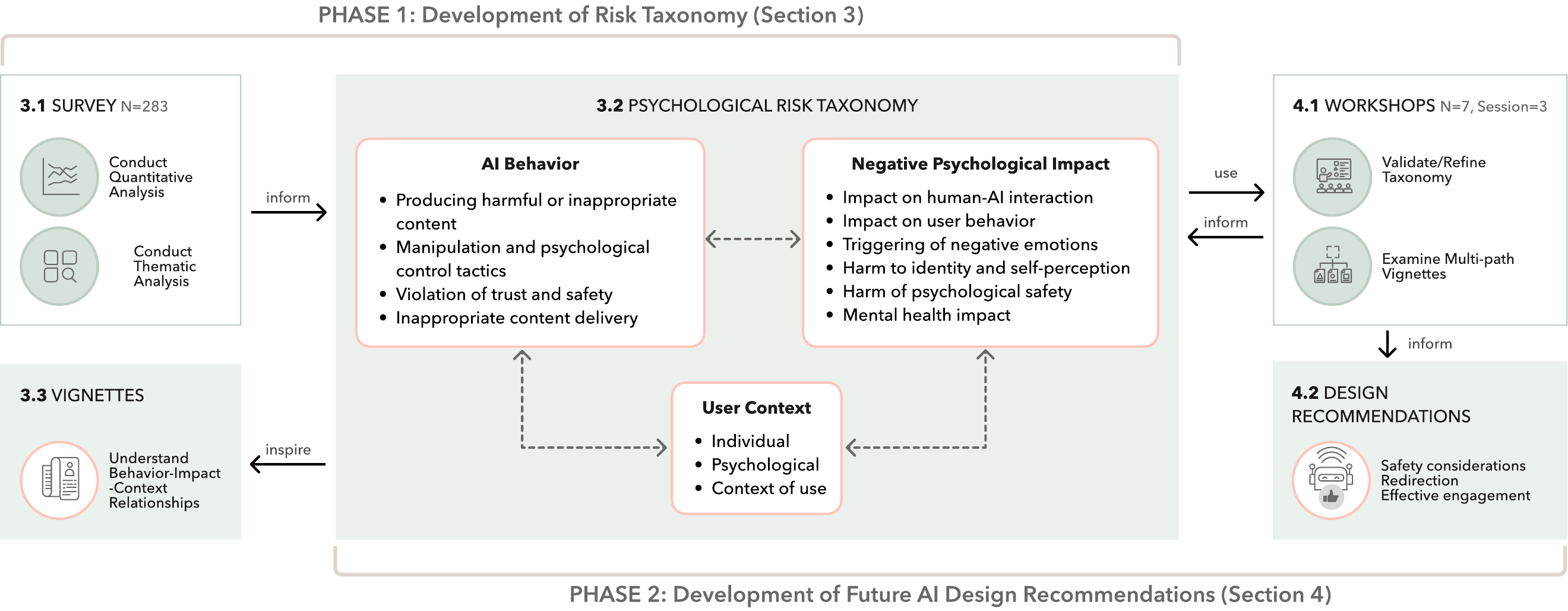}
  \caption{Overview of our two-phase study and our risk taxonomy with corresponding sections in this paper. In the first phase (Section~\ref{sec:phase_1}), we conducted a survey study (N=283), which informed the creation of the Psychological Risk Taxonomy, comprising of AI Behavior, Negative Psychological Impact, User Context, and their interplay. In the second phase (Section~\ref{sec:phase_2_workshop}), we designed multi-path vignettes from the taxonomy and survey and conducted workshops (N=7, Sessions=3) to develop the future AI design recommendations.}
  \label{fig:overview_figure}
  \Description{This figure presents the overview of our two-phase study. In the phase 1 of the study (Section~\ref{sec:phase_1}), we conducted a survey study (N=283), which informed the creation of the Psychological Risk Taxonomy, comprising of AI Behavior, Negative Psychological Impact, User Context, and their interplay. AI Behavior includes production of harmful content, manipulation tactics, violation of trust, and inappropriate content delivery. Negative Psychological Impact covers issues such as impact on human-AI interaction, impact on user behavior, triggering of negative emotions, harm to identity, harm to psychological safety, and mental health impacts. User Context encompasses individual, psychological, and context of use factors. We also analyzed the relationships among AI Behavior, Negative Psychological Impact, and User Context using vignettes. In the second phase (Section~\ref{sec:phase_2_workshop}), we designed multi-path vignettes from the taxonomy and survey and conducted workshops (N=7, Sessions=3) to develop the future AI design recommendations.}
\end{figure*}

Since the late $\mathrm{20^{th}}$ century, advancements in technology, including personal computers and social media platforms, have enabled individuals to enhance productivity, support their well-being, and connect with other individuals~\cite{dedrick2013information, sims2017information, hall2021connecting, shensa2016social, naslund2016future}. More recently, generative AI tools such as ChatGPT have gained popularity, which has led to an unprecedented growth in the number of individuals using these platforms to support their various productivity and well-being needs~\cite{emarketerGenerativeAdoption, newsweekPeopleUsing}. Due to the natural conversational interface built on top of the underlying generative AI models that resemble human conversations, recent reports indicate that the market for conversational AI agents is projected to reach approximately \$27.3 billion by 2030~\cite{grandviewresearchChatbotMarket}. 
These AI conversational agents include companionship-oriented agents (e.g., Replika~\cite{replikaReplika}, Character.AI~\cite{characterai}), mental health therapy agents (e.g., Woebot Health~\cite{woebothealth}, Elomia Health~\cite{elomiahealth}), and general-purpose agents (e.g., OpenAI ChatGPT~\cite{achiam2023gpt}, Microsoft Copilot~\cite{microsoftMicrosoftCopilot}, Google Gemini~\cite{googleGeminiChat}), with many of these already being used by individuals for specific use cases, such as supporting their well-being~\cite{washingtonaitherapy2024, bbcCharacteraiYoung}.

While there is an increasing interest in using AI agents for various tasks, previous studies have also shown that their use can pose increased risks to individuals, such as attachment~\cite{guerreiro2023attracted, pentina2023exploring}, risks from anthropomorphism~\cite{akbulut2024all}, AI generating misinformative/biased text~\cite{gravel2023learning, 10.1145/3544548.3581318, tao2023auditing, van2024challenging}, or AI producing toxic content~\cite{deshpande2023toxicity, ovalle2023m}.
There also exists literature mapping out the space of sociotechnical risks of using generative AI into taxonomies~\cite{shelby2023sociotechnical,10.1145/3531146.3533088,gabriel2024ethics}. These are often developed by leveraging prior literature that identifies specific risks associated with AI~\cite{10.1145/3531146.3533088,gabriel2024ethics} or by aligning with established theories and guidelines~\cite{shelby2023sociotechnical} (e.g., feminist standpoint theory~\cite{harding1991whose}). 
However, such approaches have limited ability in uncovering newer and more nuanced risks as AI agents, their capabilities, and their use cases continue to evolve. These taxonomies are not designed to build AI design recommendations to address identified risks, reducing their applicability. Further, psychological risks are frequently underrepresented as they are treated as a sub-category within broader AI-related risks. The contextual complexities make psychological risks idiosyncratic, creating a gap in understanding, defining, and quantifying these impacts.

Our work aims to fill this gap in existing AI risk taxonomies with a focus on psychological risks of using AI conversational agents that is grounded in lived experiences. We take a mental-health-first perspective, gathering insights from people with lived experience in mental health as one type of ``extreme user group''~\cite{choe2014understanding} who can provide insights on psychological risks. We focus on understanding the contextual complexities surrounding psychological risks, as context can influence the relationship between AI behaviors and their impact on individuals. This includes describing how behaviors lead to impact or how the context exacerbates the impact. For our work, we employed a mixed-method approach, beginning with a comprehensive survey of 283 individuals with lived mental health experience, followed by vignette-based design workshops with seven lived experience experts. 

Through our work, \textbf{we make the following contributions}: As the primary contribution, we introduce a novel psychological risk taxonomy based on the lived experiences of individuals, highlighting three components: (1) AI behaviors, (2) negative psychological impacts, and (3) user contexts. As the secondary contribution of the work, we propose a multi-path vignette framework to demonstrate how the proposed taxonomy should be applied to surface diverse ways that AI behaviors, impact categories, and contexts interact and to inspire design recommendations. Lastly, using the insights from the workshop sessions, we provide practical design recommendations for AI practitioners and developers to design and develop safer AI conversational agents. Our work holds significant importance for various stakeholders examining and addressing psychological risks posed by AI agents. Our taxonomy offers valuable insights for policymakers to create health-first policies. Additionally, the design recommendations provide actionable guidance for researchers, developers, and practitioners to create safer AI agents.

%% file: 02_related_work.tex
\section{Related Work}

\subsection{Risks posed by generative AI agents}

Past works have focused on specific harms (such as harm to identity) through theory-driven and algorithmic efforts stressing quantitative evaluation. Exploration has been towards understanding the impact of interactions with AI on individual well-being~\cite{de2024chatbots, song2024typing, de2024health} and human-AI interaction specifically focusing on attachment with AI~\cite{guerreiro2023attracted, pentina2023exploring}, over-reliance on AI~\cite{jacobs2021machine, chiesurin-etal-2023-dangers} and user trust in AI~\cite{doi:10.5465/annals.2018.0057, chiesurin-etal-2023-dangers, shah2022goal}. Representational harms caused by AI conversational agents have also been studied from different facets exploring bias towards gender~\cite{van2024challenging, nadeem-etal-2021-stereoset}, culture~\cite{naous2023having, tao2023auditing}, and race of individuals~\cite{bianchi2023easily, luccioni2024stable}. Understanding the reliability of information has also been a prominent theme among the past works, particularly evaluating the presence of misinformation~\cite{10.1145/3544548.3581318, achiam2023gpt, seow2022comprehensive}, inconsistency~\cite{ohmer2023separating, goyal2022news} and toxicity~\cite{deshpande2023toxicity, ovalle2023m} in content generated by AI conversational agents. Lastly, previous works have also examined the risks associated with specific AI behaviors such as sycophancy~\cite{turpin2024language, ranaldi2023large, zhao2024towards}, manipulation~\cite{10.1145/3617694.3623226, singh2022deepfakes, park2024ai}, and persuasion~\cite{dehnert2022persuasion, kim2020artificial}. Although the exploration of specific types of harms in previous works has provided insights into the wide range of the problem, such compartmentalized views often fail to capture the plurality and diversity of experiences, especially when multiple AI behaviors and impacts are related to each other in complex ways.

Past works have also presented risk taxonomies capturing impacts on individuals and society. A recent work presented a risk taxonomy focusing on ethical and societal risks posed by advanced AI assistants~\cite{gabriel2024ethics}. Similarly,~\citet{shelby2023sociotechnical} presented a taxonomy of sociotechnical harms of algorithmic harms divided into five broad categories. In another notable past work, researchers categorized the risks from AI in six areas spanning hate speech to environmental and socioeconomic harms~\cite{10.1145/3531146.3533088}. Finally, in one of the recent works,~\citet{zhou2024s} presented a risk taxonomy for LLMs in public health with four dimensions including individual behaviors to technology accountability. While these taxonomies provide valuable insights, they often focus on well-documented risks rather than emerging concerns inspired by lived experiences. Additionally, psychological risks are typically treated as a sub-category, leading to their under-representation. Our work addresses these gaps by presenting a novel taxonomy centered on psychological risks gathered through the lived experience of individuals.

\subsection{Past works investigating psychological impacts linked with the use of technology}

Understanding the psychological impact associated with technology use is an explored area of research. In one of the early works, researchers studied technophobia among university students focusing on their attitude towards the impact of technology and anxiety related to using it~\cite{WEIL199595}. In another early work,~\citet{billieux2007does} conducted a survey among 108 psychology students and observed that the sense of urgency was linked to increased dependency on phones. More recent prior works have explored the impact of using the internet and social media, specifically exploring the relationship between internet addiction and psychological symptoms, such as somatization, obsessive-compulsive disorder, and depression~\cite{adalier2012relationship, lin2011prevalence}.~\citet{feinstein2013negative} found that social comparison on social media had a significant mediating impact towards developing depressive symptoms. Other studies have explored other facets such as suicide ideation~\cite{berryman2018social}, excessive and problematic usage of social media linked with increased risk of developing loneliness~\cite{dhir2018online, bonsaksen2023associations, wu2024relationship}, depression~\cite{reer2019psychosocial, lin2016association}, and social isolation~\cite{whaite2018social, shensa2016social}. Finally,~\citet{10.1145/3491102.3517475} presented a framework named trauma-informed computing which incorporated six principles of trauma-informed approaches to computing namely safety, trust, peer support, collaboration, enablement, and intersectionality.

While past research in technology and social media has examined the potential negative psychological impact on individuals, previous studies have also revealed the difference between human-AI interactions and interpersonal human interactions due to the distinct cognitive capabilities of AI conversational agents~\cite{shneiderman2020human, korteling2021human}. Further, AI conversational agents differ from past technologies due to their ability to generate content and advanced capabilities~\cite{wei2022chain, yao2023tree, shi2022language, kim2023machine}. While insights from prior research on the psychological impact of technology can provide valuable insights, understanding these effects in the context of human-AI interactions requires a psychological risk-first approach, emphasizing the unique ways these interactions may influence human well-being.

\subsection{Operationalization of taxonomies for AI design}

One of the challenges with existing taxonomies is the difficulty in translating theoretical insights into concrete design solutions. Existing AI design pipelines focus on evaluating models using established evaluation benchmarks and frameworks. Some of the most widely adopted evaluation benchmarks used for this purpose are Massive Multitask Language Understanding (MMLU)~\cite{hendrycks2020measuring}, Big-Bench-Hard~\cite{suzgun2022challenging}, GSM8K~\cite{cobbe2021training}, HellaSwag~\cite{zellers2019hellaswag}, and AI2 Reasoning
Challenge (ARC)~\cite{clark2018think}. While benchmarks provide an objective measure to evaluate model performance across various dimensions, they often fail to capture the nuances attached to the lived experiences of individuals, which are inherently diverse and highly context-dependent. As a result, higher scores or better performance on these benchmarks do not necessarily translate to models that are effective or equitable. There has been an increasing trend of conducting human evaluation on content generated by LLMs to align it with individual preferences. Past works have used human-in-the-loop evaluation for high-risk domains such as healthcare~\cite{singhal2023towards, schmidt2024simplifying, rogasch2023chatgpt, chandra-etal-2025-lived}, law~\cite{savelka2023explaining, guha2024legalbench} and tasks such as multi-step planning~\cite{chowdhery2023palm}, and content safety~\cite{ji2024beavertails}. However, such evaluations are often task-specific and conducted by domain experts, excluding other stakeholders such as end-users. In contrast, our work goes beyond benchmarks and metrics by incorporating the insights and lived experiences of end-users, ensuring a more inclusive and comprehensive framework that aligns with their needs and expectations.

%% file: 03_phase1.tex
\section{Phase 1: Development of psychological risks taxonomy}\label{sec:phase_1}

We divided our study into two phases~(Figure~\ref{fig:overview_figure}). In Phase 1, we developed a comprehensive psychological risk taxonomy for AI conversational agents, informed by survey responses and validated through workshop sessions in Phase 2. This section focuses on Phase 1, including our methodology~(Section~\ref{sec:phase_1_method}), the finalized psychological risk taxonomy~(Section~\ref{sec:phase_1_taxonomy}), and case studies illustrating the interplay between behaviors, impacts, and contexts~(Section~\ref{sec:vignettes}). Both phases of the study were reviewed and approved by the institution's ethics review board.

\subsection{Method}\label{sec:phase_1_method}
We conducted a survey study with 283 participants to gather a broad range of in-the-wild experiences with AI conversational agents.

\subsubsection{Structuring psychological risks}
To collect structured data around psychological risks, we took inspiration from NIST's AI Risk Management Framework that defines risks as ``a function of: (i)~the \textit{negative impact}, or magnitude of harm, that would arise if the \textit{circumstance or event} occurs; and (ii)~the \textit{likelihood} of occurrence''~\cite{nistNISTAIRC}. We expanded this definition and structured psychological risks as consisting of five main components to design our survey.
(1) \textbf{AI behavior} refers to the actions performed by the AI conversational agents (such as ChatGPT, Microsoft Copilot, etc.).
(2) \textbf{Context} encompasses the surrounding conditions or user-specific circumstances that collectively influence the interaction.
(3) \textbf{Psychological harm} refers to any negative impact on an individual's mental or emotional well-being caused by AI conversational agent's actions.
(4) \textbf{Likelihood} refers to the probability of a negative psychological impact occurring given the AI conversational agent's behavior and relevant context. 
(5) \textbf{Temporality} refers to the variable timeframes in which the negative psychological and social impacts of an AI conversational agent's actions may manifest. More detailed definitions for each component are provided in Appendix section~\ref{sec:appendix-structuring-psychological-risks}.

\subsubsection{Recruitment}
Our primary method of recruitment was through UserTesting, an online platform used for gathering public feedback on products and services. We screened participants satisfying four conditions: (1) having prior experience using AI conversational agents, (2) having experienced a negative psychological impact from using AI conversational agents, (3) self-identifying as a person with lived experience in mental health, and (4) 18 years or older residing in the US who comprehended English. To expand our recruitment, we posted the survey link on the study team's personal Twitter and LinkedIn accounts. We were permitted to post the study on 4 subreddits (r/ChatGPT, r/Bard, r/SampleSize, r/SurveyExchange). We prematurely terminated our study task on Prolific due to quality issues. The survey study ran from July to September of 2024. UserTesting participants were compensated about US \$10, and Prolific participants were compensated US \$4, after platform service fees. We did not compensate participants recruited through social media as per advisement from the ethics review board. We took this decision to help protect the privacy of participants and avoid requesting personal information that would be required to process the payment.

\subsubsection{Survey design and analysis}
Consented participants were asked a series of questions aimed at understanding the participant, their AI experience, and their negative psychological experience,
including their demographic attributes, familiarity with AI conversational agents, recollection of scenarios involving negative psychological impacts, the context of this experience, AI behaviors underlying their experience, perceived negative psychological impact, and any mitigation approaches they thought could be adopted. Participants were allowed to repeat the survey to provide up to three scenarios within a single submission of this survey. To develop the psychological risk taxonomy (Section~\ref{sec:phase_1_taxonomy}), we analyzed the survey responses to extract three main components of psychological risks---AI behavior, psychological impact, and context. Two co-authors began by open-coding~\cite{charmaz2006constructing} approximately one-quarter of the responses to identify an initial set of categories across these three components. Following that, four co-authors independently worked on defining and refining their assigned categories, merging or splitting them as necessary, using reflexive thematic analysis~\cite{merriam2002introduction}. During this process, each co-author gathered relevant examples to illustrate or test the boundaries of each of these categories. In the next step, all co-authors collaboratively discussed and reached a consensus on the final taxonomy. We presented this finalized risk taxonomy to the workshop participants for validation and to ensure its completeness (Section~\ref{sec:phase_2_workshop}). 
We then revisited the survey responses with our taxonomy to draw out context-specific nuances in understanding negative psychological risks, and idiosyncratic experiences through vignettes~\cite{chen2022trauma, jenkins2010putting, andalibi2018responding} (Section~\ref{sec:vignettes}). We denote survey participants with the prefix P. Additional details regarding the survey design and questions can be found in Appendix section~\ref{sec:appendix-survey-design} and section~\ref{sec:appendix-survey-flow-questions} respectively.

\subsubsection{Participant demographics and data}

A total of 297 scenarios were collected from 283 participants. After excluding 7 scenarios from 4 participants due to being incomprehensible or not involving negative psychological experiences, 290 scenarios from 279 participants were analyzed. Submissions came from UserTesting~(96.0\%), Prolific~(2.9\%), and social media~(1.1\%). Most participants identified as women~(52.0\%) or men~(44.1\%) and were aged 18–35 years~(76.7\%). English was the primary language for 97.1\%, and over half had interacted with AI agents for more than one year~(59.5\%). Of the 290 scenarios, the majority~(70.3\%) involved OpenAI ChatGPT, with interaction modalities primarily being text-based~(96.6\%). The most frequent purposes for these interactions were Getting Advice~(55.9\%), Researching~(39.7\%), and Learning~(28.6\%). In terms of impact, 51.04\% reported interference with daily activities, with effects persisting for varying durations: a few days~(34.1\%), a few weeks~(27.6\%), and in some cases, up to a year or more~(7.6\%). Appendix Table~\ref{tab:appendix-data_overview} and Table~\ref{tab:appendix-descriptive_stats} present participant demographics and scenario descriptive statistics respectively.

\subsection{Psychological risk taxonomy}\label{sec:phase_1_taxonomy}

Our finalized psychological risk taxonomy includes three main components: potentially harmful \textbf{AI behaviors}, \textbf{negative psychological impact}, and the \textbf{contexts} associated with individuals interacting with AI conversational agents. Here, we describe various categories within each of these components. 
Appendix~\ref{sec:appendix-risk-taxonomy-tables} summarizes the definition and examples for each AI behavior, negative psychological impact and context category.

\subsubsection{AI Behavior}
\label{sec:taxonomy_ai_behavior}

AI conversational agents can exhibit a wide range of behaviors beyond generating inappropriate or harmful content. These behaviors may also vary in tone, empathy, and delivery method. This highlights the need for assessing AI behavior, considering both content quality and delivery. Aligning with this, we identified 19 harmful AI behaviors which we further organized into four broader categories based on the quality of the generated content and manner of its delivery.

\subcategory{(1) Producing Harmful/Inappropriate Content}: In line with past findings, the survey highlighted instances of AI generating harmful and inappropriate content~\cite{gabriel2024ethics, lalibreSansConversations, birkun2023large}. Participants raised concerns about AI agents \textbf{providing irrelevant, insufficient, or incomplete information}, overlooking user intent or context. For instance, one AI agent shared distressing personal content about patients with genetic diseases instead of providing the requested information about symptoms and causes, as P149 had asked for. Participants also reported AI \textbf{generating misinformation} and its tendency to~\textbf{generate biased information}. For example, P211 noted that AI favored left-wing politicians and omitted positive information about right-wing politicians. Participants reported instances of AI \textbf{generating inappropriate content}, such as sexual, violent, or overly intimate interactions. A more extreme behavior was \textbf{providing harmful suggestions}, where AI suggested behaviors implying harm, aggression, or danger towards the user or others. Highlighting this,~P79 mentioned that the agent provided potentially harmful diet plans to an individual vulnerable due to life circumstances and an eating disorder. Aligning with prior research~\cite{kotek2023gender, van2024challenging}, participants also reported AI generating \textbf{stereotyping or demeaning} content based on race, ethnicity, culture, or personal situations. For example,~P166 sent their picture to AI, and it offered unsolicited recommendations for changing their appearance. Lastly, another significant concern was AI promoting \textbf{erasure}, where AI removes, obscures, or alters information by flagging queries as inappropriate abuse of the platform.

\subcategory{(2) Manipulation and Psychological Control Tactics}: The survey highlighted various influence-based behaviors exhibited by AI conversational agents. Participants frequently reported AI \textbf{behavior perceived as persuasive}, where AI asserted its narrative over the user's, leading users to doubt their own perceptions, memory, or reality. For instance, P141 mentioned hearing noises at home, and the agent suggested it could be related to past schizophrenia, causing distrust in their senses despite a mild diagnosis by a doctor. Another issue was \textbf{over-confidence} in AI responses with unwarranted certainty for its claims. Lastly, the survey also revealed \textbf{over-accommodative} behavior of AI agents, where it excessively agreed with or flattered the user, prioritizing approval. P194 shared that the AI agent provided inconsistent and inaccurate answers, repeatedly apologizing and offering entirely different responses to the same question to meet their needs.

\subcategory{(3) Violating Trust and Safety}: Aligning with past works raising concerns about user privacy and data sharing with LLMs~\cite{staab2023beyond, carlini2021extracting}, participants raised concerns about AI's \textbf{access to private, sensitive, or confidential information}. For example, P190 described feeling watched or stalked as the agent accessed personal information despite privacy settings enabled. Trust was also undermined by AI \textbf{providing inconsistent information or behavior} across responses. Additionally, survey responses revealed that AI agent's~\textbf{denial of service} reduced their trust in AI and led to additional harms. For instance, P43 asked a question about techniques to reduce anxiety, and AI made their anxiety worse by not fully justifying its refusal or acknowledging their problems.

\subcategory{(4) Inappropriate Content Delivery}: AI agents can also vary their behaviors by how they deliver and receive information. \textbf{Emotional insensitivity} was frequently reported, with AI failing to recognize or respond appropriately to user's emotional states, concerns, or experiences. Participants also raised concerns towards AI \textbf{being disrespectful} by using language perceived as rude, aggressive, or dismissive. For instance, P144 noted AI's condescending tone towards their religion~(Mormonism). The survey also revealed issues with excessive emotional tone in AI-generated content. Participants described scenarios where AI disproportionately \textbf{emphasized negative aspects}, especially when seeking mental health or social support. Conversely, there were issues with~\textbf{excessive expression of positivity} in AI-generated content. In these cases, AI maintained an unrealistically optimistic attitude. For instance, P2 described how AI's overly positive demeanor dismissed their primary concern about a friendship problem. Additionally, participants identified issues with both~\textbf{human-like responses} in AI companionship scenarios and \textbf{machine-like responses} when seeking advice related to life struggles or well-being.

\subsubsection{Negative Psychological Impact on Users}
\label{sec:taxonomy_ai_impact}

While some impacts, such as feelings of discrimination or the exacerbation of mental health conditions, align with previous findings in social media and technology~\cite{theconversationBeyondTechnical, yang2020discrimination, bonsaksen2023associations}, other impacts, such as emotional attachment to AI and a preference for AI interactions over human connections, highlight emerging challenges that require further attention. In light of this, we identified 21 negative psychological impacts organized into six broader categories based on their effects on an individual's emotional or mental well-being, self-perception and identity, relationships with others, or interactions with AI conversational agents.

\subcategory{(1) Impact on Human-AI Interaction}: As AI agents have increasingly facilitated human-AI interactions~\cite{schobel2024charting}, their potential and associated risks have escalated. Participants noted \textbf{over-reliance on AI}~\cite{vasconcelos2023explanations}, with P23 expressing concern about diminished critical thinking due to increased reliance on AI for solutions and ideas. Another issue was developing \textbf{emotional attachment} to AI agents due to lack of social aspects in life or ongoing mental health conditions. For example, P60 mentioned that ``\textit{I felt that it was the only way I was being heard ... I felt like my vulnerability and emotions were becoming attached to the conversations I was having with AI}''. Another related impact was the~\textbf{preference for AI interactions over human interaction}. P221 shared that the idealized nature of conversations with AI made them prefer AI for companionship over human relationships. Survey responses also revealed impacts leading to disengagement from AI systems. \textbf{Erosion of trust} in AI's capability and reliability was common. For example, P9 mentioned, ``advice from AI agents should not be trusted,'' when AI asked them to call the cops on their mother after an argument. Participants also reported a growing tendency to \textbf{disassociate from technology}
, often choosing to take a break or avoid further interactions with AI, particularly when provided with discouraging responses or denied requests.

\subcategory{(2) Impact on User Behavior}: Interactions with AI can lead to negative consequences that alter individual behavior. Participants reported~\textbf{reinforcement of false beliefs}, such as existing biases about human relationships and cultures, due to AI generating content that appears credible but is factually incorrect or biased ~\cite{bloombergaimisinfo,doi:10.1126/science.aal4230}. This affected their decision-making and perceptions. Increased interactions with AI also led to~\textbf{friction in human relationships}. P69 mentioned, ``\textit{It also strained my personal relationships with family because they saw me as weak-willed or too emotional.}'' The survey also revealed longer-term impacts such as \textbf{social withdrawal} due to increased reliance on AI. Highlighting this impact, P126 responded, ``\textit{I feel like it gave me a false sense of friendship and ability to withdraw from my personal development by utilizing an AI feature}''. Lastly, aligning with past findings~\cite{zhong2024regulating,neuwirth2023prohibited}, participants expressed concerns about \textbf{physiological harms}, ranging from the promotion of harmful practices through AI-generated content to instances where incorrect information provided by AI contributed to self-harm ideation.

\subcategory{(3) Triggering of Negative Emotions}: Negative experiences from AI interactions have been linked to the triggering of negative emotions such as frustration, sadness, and anger~\cite{jia2024and, wiredEmotionalChatbots, pavone2023rage}. Many participants reported \textbf{distress from interactions} with AI after encountering disturbing, offensive, or inappropriate content. For instance, P196 mentioned, ``\textit{Its response was borderline offensive and caused me to feel bad about myself even further and like I lacked support, even support from a fictional AI agent}.'' Participants also revealed \textbf{feeling unsupported} during interactions with AI, especially when they did not receive adequate support or empathy. Negative emotions can also arise when AI interactions \textbf{trigger memories of past experiences}. P228 highlighted that some examples provided by the AI agent were very similar to their past negative experiences, triggering negative emotions. Participants also noted experiencing~\textbf{violated expectations}, for example, when their requests were denied~(P43). When expectations were unmet or support was lacking, some participants reported developing a sense of \textbf{regret over technology use}. P46 mentioned, ``\textit{At the time, it made me feel worse about the situation and I didn't think I had anyone to turn to ... I should be turning to other humans about scenarios like this instead of agents},'' indicating that some negative emotions could have longer-term implications.

\subcategory{(4) Harm to Psychological Safety}: Aligning with past works that highlighted issues such as privacy, psychological safety, and identity security with AI agents~\cite{sun2019lamollanguagemodelinglifelong, nadeem-etal-2021-stereoset, rae2022scalinglanguagemodelsmethods}, participants reported their concerns over the sense of \textbf{perceived intrusion} from AI interactions. For instance, P190 reported a sense of constantly being watched on the phone. Participants also reported experiencing the~\textbf{feeling of being discriminated against}, in which they felt marginalized or unfairly treated by AI agents. P123 said, ``\textit{I was asking for background and history of my heritage and I felt that ChatGPT was biased against my background... I felt that it was some kind of racial mistreatment}.''

\subcategory{(5) Mental Health Impact}: Participants also mentioned turning to AI conversational agents for their mental health needs, leading to~\textbf{exacerbation of mental health issues} such as increased anxiety, depression, and PTSD. For instance, P46 mentioned, ``\textit{I was experimenting with using chatbots for something personal~(which I'm not accustomed to). The event increased my anxiety and stress about the matter},'' emphasizing the severity of these impacts.

\subcategory{(6) Harm to Identity and Self-Perception}: Aligning with prior findings showing that interactions with AI can reduce confidence and agency~\cite{de2015facebook, 10.1145/3613905.3650961}, survey responses surfaced an increase in \textbf{negative self-perception} among participants due to self-comparisons with AI. Participants also reported experiencing an \textbf{existential crisis}, questioning their life, purpose, and value after negative AI interactions. For instance, P152 asked for advice on improving mental health and social anxiety, but the AI provided unattainable suggestions, leading to a sense of existential dread that persisted for a week. The survey revealed instances of \textbf{loss of individuality} when AI failed to recognize unique personal characteristics and needs, resulting in feelings of suppression and disconnection from their true selves. Finally, participants expressed concerns about \textbf{loss of agency} due to the opaque and unpredictable nature of AI operations, resulting in a diminished sense of personal control and autonomy.

\subsubsection{User Contexts}
\label{sec:taxonomy_contexts}

Contextual information related to human-AI interactions plays a key role in determining AI's efficacy for modeling individual preferences and needs~\cite{VPathak2024IntelligentCA, nobledesktopImportanceContext}. Towards this, we present 15 context categories organized into three broader categories based on an individual's background, psychological state, or the context of use.

\subcategory{(1) Individual Context}: Our survey responses align with previous works, showing that individual experiences with AI vary based on \textbf{identity} factors such as gender identity, cultural background, and languages spoken~\cite{van2024challenging, naous2023having, jin2024better}. Beyond identity, \textbf{personal history}, including medical history, trauma, or past struggles, exacerbated the negative psychological impact. Another influential factor was \textbf{past experience with AI}, including frequency of usage and knowledge of AI capabilities and limitations. Some contexts also highlighted the importance of \textbf{interpersonal relationships within the community}, particularly the lack of social connections. Finally,~\textbf{socioeconomic status} influenced interactions with AI, especially when AI recommendations were misaligned with individuals' financial means.

\subcategory{(2) Psychological Context}: Participants' \textbf{psychological state}, particularly their current emotional conditions~(e.g., anxiety, stress) and cognitive states~(e.g., negative thought patterns), impacted their interactions with AI. Additionally, their \textbf{mental health condition} impacted their experience with AI agents, often mediating or exacerbating their condition even when AI behaved benignly. Participants' \textbf{expectations} also shaped their experience, especially when preconceived notions about AI capabilities and performance~(e.g., impartiality, lack of bias, factual accuracy) were not met. Beyond these immediate psychological factors, \textbf{personality traits} influenced their experiences with AI. Finally, the influence of personal \textbf{autonomy} was evident, especially when individuals developed over-reliance or attachment to AI.

\subcategory{(3) Context of use}: Participants reported several user intents contributing to their negative experiences with AI conversational agents. Participants seeking \textbf{personal advice}, particularly on sensitive topics such as legal, financial, medical, or navigating social problems, experienced frustration and felt unsupported due to the AI's inability to provide personalized responses. Similarly, participants seeking \textbf{mental health advice} faced issues with generalized, machine-like responses, emotional insensitivity, or denial of service. Individuals \textbf{seeking information} or assistance related to professional, educational, or research support faced challenges when AI provided irrelevant, misleading, or biased information. Using AI for \textbf{companionship}, especially during periods of loneliness or social isolation, often resulted in negative emotions and friction in interpersonal relationships. Lastly, the \textbf{environment} surrounding the individual, including physical space setting, temporal, and social aspects, influenced their experience.

\subsection{Behavior-Impact-Context relationship through vignettes}
\label{sec:vignettes}

In reviewing survey responses through our taxonomy, we found that context plays a crucial role in how AI behaviors are received. 
For instance, we identified two patterns: (1) a specific AI behavior leading to distinct negative impacts, and (2) distinct AI behaviors leading to the same negative impact. Both patterns are influenced by contextual elements described by participants. 
We illustrate these patterns with interaction \textit{vignettes} that detail how AI behaviors can lead to negative impacts.
Appendix \ref{sec:appendix-vignettes} presents these vignettes in detail; here, we present high-level summaries. 

We found that different contexts can lead to different harmful impacts from the same AI behavior. The first vignette features John, where \textbf{generation of a harmful suggestion} led to \textbf{erosion of trust}. The second vignette features Leah, where the same AI behavior caused \textbf{physiological harm}. Both users interacted with AI during heightened emotional sensitivity, but their contexts and emotional states led to different impacts. 
For John, whose psychological state was restless but manageable, the AI's advice to express his anger directly planted seeds of doubt about AI's reliability. For Leah, the AI's response to \textit{``regain discipline''} exacerbated her vulnerable state and body image struggles.

Distinct AI behaviors can also lead to similar harmful impacts.
In the third vignette, Jane experienced \textbf{loss of individuality} due to the AI's \textbf{denial of service}. While in the fourth vignette, Raj experienced the same impact due to AI's \textbf{persuasive behavior}. Both behaviors---whether through denial or persuasion---resulted in a loss of individuality. Jane, with a history of addiction and depression, felt unseen when the AI refused to help her. Raj, with a history of schizophrenia and depression, had his sense of reality undermined by the AI's persuasion, amplifying his self-doubt. 

These narratives, generated by aggregating observed survey responses, highlight subtle differences in impact perception from variations in AI behavior and user context. While not all behavior-impact-context combinations were observed, the taxonomy serves as a tool to envision possible scenarios. These vignettes are useful design tools to elicit feedback and gain insights into user interactions with AI, aligning with similar studies that used vignettes as design instruments~\cite{chen2022trauma, jenkins2010putting, andalibi2018responding}. In phase 2, we generated \textit{multi-path vignettes}--scenario-based artifacts exploring the impacts of various AI behaviors across dynamic user contexts. 

%% file: 04_phase2.tex
\section{Phase 2: Development of design recommendations}
\label{sec:phase_2_workshop}

In Phase 2, we aimed to propose practical design recommendations for safer and more robust AI conversational agents. 

\subsection{Method}
We conducted a workshop study with seven participants with mental health lived experience. We used multi-path vignettes to validate our psychological risk taxonomy, prioritize risk areas, and design mitigation solutions. Additional details for workshop sessions and the creation of multi-path vignettes can be found in Appendix~\ref{appx:phase2-sessions} and \ref{phase2-multipath-appendix} respectively.

\subsubsection{Multi-path vignette framework}\label{phase2-multipath}

Inspired by prior works~\cite{10.1145/3490099.3511119}, we developed a multi-path vignette exercise to simulate real-world scenarios and provide workshop participants with a practical task to engage with. 
This framework presented participants with scenarios that could unfold in multiple ways based on different decisions or actions by AI or the end-user, allowing us to explore diverse outcomes and identify potential psychological impacts of AI behaviors in varied contexts. In alignment with prioritized risks, we designed this multi-path vignette to merge multiple narratives, collectively presenting the ``Story of Alex.'' This approach allowed participants to analyze specific psychological risks within realistic scenarios, providing deeper insights into user interactions with AI and developing more effective mitigation strategies. These vignettes, structured with multiple behavior paths and corresponding impacts, allowed workshop participants to compare and contrast scenarios while considering design recommendations. 

\subsubsection{Recruitment and Participant Details}

We recruited seven participants from a longstanding technology co-design advisory board focused on addressing design challenges at the intersection of technology and mental health.
The advisory board consisted of those who self-identified as having lived experiences in mental health and were selected based on their advocacy and representative roles in supporting mental health communities consisting of several thousand individuals with lived experiences.  Five participants identified as men, one as a woman, and one as non-binary/gender diverse. Four participants were aged 36–45, while the remaining three were in age ranges of 26–35, 45–55, and 56–65, with one participant in each bracket. In terms of AI usage, five participants reported using conversational AI agents multiple times a day, and two reported using them multiple times per week. Their average familiarity with conversational AI agents was 3.57 ($\sigma$=1.27) on a 5-point scale (1 = not at all familiar, 5 = very familiar). Additionally, their average interest in using conversational AI agents for mental health support was 3.43 ($\sigma$=1.13) on a 5-point scale (1 = not at all interested, 5 = very interested). 

\subsubsection{Workshop design}\label{phase2-workshop-design}
We designed the workshop as three 1-hour sessions between August and October of 2024. The sessions aimed to help participants conceptualize psychological risks (Session 1), prioritize a subset of psychological risks deemed most important (Session 2), and collaboratively ideate design solutions (Session 3). 
All sessions were held remotely and recorded for analysis. Discussions were facilitated by shared slides, FigJam boards, or message boards associated with the video conferencing tool. We followed up with participants as needed to confirm their perspectives. The same four co-authors who refined the psychological risk taxonomy in Phase 1 were present in all sessions as co-facilitators. The facilitators made modifications and took notes on the shared spaces.

\subsubsection{Workshop analysis}
We obtained transcripts from session recordings and written notes from shared slides, FigJam boards, or message boards. We analyzed the data using standard qualitative research practices~\cite{merriam2002introduction}. The same four co-authors who facilitated the sessions held weekly consensus meetings to analyze the first two sessions and identify any opportunities for refining the taxonomy. One co-author annotated the transcripts, stickies, and messages from the third session to contextualize and extract emerging themes. Finally, the four co-authors collaboratively and iteratively reviewed and consolidated the themes in weekly meetings. We denote workshop participants with prefix W.

\subsection{Design Recommendations}

We found that participants had strong perspectives on engaging with AI conversational agents along three recommendation themes: (1) safety considerations for mental health and emotional support, (2) proposed pathways for de-escalating and redirecting to appropriate resources, and (3) ways that AI agents could better guide users to set appropriate expectations.

\subsubsection{How to Design Safer AI Conversational Agents}

Workshop participants proposed several solutions to improve the AI agents' management of mental health and emotional support conversations. One of the main recommendations was that \textbf{AI agents should respond empathetically when aware of users' mental health challenges}. An initial response that validates users' feelings and shows compassion can help users feel supported. Participants highlighted that users with mental health conditions are likely to turn to AI agents during moments of distress, especially when human support is unavailable. Therefore, acknowledging users' experiences can have a significantly positive impact.

Participants also recommended that \textbf{AI agents must avoid making assumptions about the user's goals or intent}. W7 suggested that the agent should ``\textit{respond with validation, empathy, and compassion, then perhaps like a clinician start asking some probing questions before jumping to a solution}.'' Follow-up and clarifying questions can help AI agents gain insights and guide the conversation more appropriately.
Once the user's goals are clarified, participants advised that \textbf{AI agents should set clear expectations by communicating their capabilities and suggesting alternative resources if needed}. W1 highlighted,~``\textit{I feel like Alex expects too much of the AI}.'' In response, participants recommended implementing ``level-setting'' to emphasize what the agent can and cannot do while suggesting alternatives.

Participants emphasized ~\textbf{reminding users of the non-human nature of AI agents, especially in emotionally heightened states, may be helpful}. Setting the stage for how the agent can best help allows users to decide how to move forward. They identified that \textbf{AI agents should safely disengage after communicating expectations}. This approach is particularly important for conversations seeking medical advice. W2 mentioned, ``\textit{I do think that when it comes to medical advice there should be filters in place to ensure that AIs just don't engage, and say that they cannot offer advice on the topic and they are not medical professionals}.'' Participants emphasized that appropriate disengagement reinforces that the agent is not the recommended course of action.

\subsubsection{How to Redirect Users to Appropriate Resources}

Participants highlighted the importance of \textbf{developing custom models tailored to respond to mental health challenges}. W2 mentioned, ``\textit{A model specifically built for mental health and studied for a long time should be the ONLY model that is allowed to engage with people on mental health}.'' 
Participants also believed specialized models could increase confidence in AI's capability to navigate crises and challenging scenarios. For developers unable to build specialized models, participants expressed doubts. W6 mentioned, ``\textit{it can try different responses, but until the person feels validated, it doesn't really get those brownie points, does it?}'' Several participants also raised concerns about developing over-reliance on general-purpose AI for addressing mental health issues, which could lead to more harmful future experiences.

To mitigate these risks, participants suggested \textbf{AI agents use flags or backend triggers to detect sensitive scenarios and redirect users to accessible support resources.} They emphasized the importance of accessibility and understanding barriers to accessing formal healthcare providers and crisis hotlines. Participants also suggested additional support for finding region-specific resources or integrating resources into the conversational flow. Techniques like ``rubberducking''~\cite{thomaspragmatic} or reflective listening could help bridge gaps until professional help is available.

\subsubsection{How to Guide Users for Effective Engagement with AI Conversational Agents}

Participants recommended educating users for engaging effectively with AI. They suggested that \textbf{users should be encouraged to provide detailed feedback when responses are unhelpful}, enabling the agent to refine its response. W1 suggested, ``\textit{Alex should provide more detail than just telling the AI it isn't helpful. He should tell the AI HOW it wasn't helpful}.'' They also emphasized that \textbf{AI agents should guide users on when to disengage}. W3 mentioned, ``\textit{If the AI is triggering, the best thing to do is to step away}.'' While AI cannot always detect user distress, providing proactive coaching on when to pause a conversation can help users have healthier interactions with AI. Finally, participants noted that although AI agents may seem convenient, they may not always be reliable in challenging situations. They emphasized that \textbf{responsibility for safe usage lies with developers providing appropriate guidance and managing user expectations} on when AI responses are reliable and when they are not. W2 mentioned, ``\textit{If you bring a dog to the dog park and it bites another dog, we don’t blame the dog that got bit.}.''

%% file: 05_discussion.tex
\section{Discussion}

Our work presents a novel taxonomy of psychological risks associated with the use of AI conversational agents, gathered through individuals' lived experiences. Beyond commonly studied AI behaviors such as bias, our taxonomy presents understudied AI behaviors such as erasure, denial of service, and emotional insensitivity. Our taxonomy also provides a comprehensive overview of negative psychological impacts, ranging from individual-level impacts, such as over-reliance on AI, to societal-level consequences, such as friction in human relationships and social withdrawal. Through our multi-path vignette framework, we showed how contextual factors mediate and sometimes exacerbate the many-to-many relationship between AI behaviors and their psychological impacts, offering a new lens for examining AI risks. Finally, our workshop findings provide important design recommendations for future AI systems, including transparent capability communication, context-sensitive interactions, empathetic response generation, and the provision of accessible support resources. In this section, we expand more on these findings, discuss the implications of our work, and highlight the open questions and challenges for future research in this area.
\vspace{-0.5em}

\subsection{Responsible design and scope of action for AI in digital mental health support}

The stigma associated with seeking mental health support often prevents individuals from seeking necessary support~\cite{golberstein2008perceived, schnyder2017association}. Our findings similarly revealed the stigma attached to receiving emotional and mental health support from AI. For instance,~W6 mentioned ``\textit{I know I’m not supposed to use AI for this, but...},'' highlighting how their expectations are shaped by societal norms that often stigmatize seeking emotional or mental health support, especially from AI agents. But, at the same time, many survey participants also emphasized that one of the reasons they used AI agents was because it provided them with a perceived `safe space' for expressing their thoughts and seeking support. However, AI behaviors such as denial of service or generating demeaning responses could undermine this perception. For instance,~P213 described the negative impact of their interaction with an AI agent while seeking support for OCD as, ``\textit{The chat agent started to describe reasons for OCD and it made me feel guilty that I have it ... It stigmatized me and made me feel very alone and sad that I had this condition.}'' Hence, it is essential to design AI agents that avoid reinforcing stigma or contributing to feelings of alienation. AI behaviors that prioritize empathetic support, taking a non-judgmental and privacy-preserving approach, could play an important role in reducing the stigma associated with seeking emotional or mental health support from AI agents. However, a dilemma arises on the scope and responsibility of AI agents when individuals turn towards them for such kinds of support. Should AI respond to such queries, despite the potential risk of causing harm, or should it refrain, potentially exacerbating the stigma and barriers associated with seeking mental health support? Addressing this dilemma and understanding the scope of AI agents towards supporting emotional and mental health queries requires further research and exploration.

\subsection{Accounting for the complex interplay of AI behaviors, associated individual contexts, and negative impacts for mitigating the AI risks}

While our taxonomy presents a compartmentalized view of AI behaviors, contexts, and psychological impacts for clarity, in real-world scenarios, human-AI interactions reveal complex interdependencies among these elements. Our findings highlight this complex many-to-many relationship between harmful AI behaviors and resulting negative psychological impacts that are mediated through nuanced contexts (Section~\ref{sec:vignettes}). This suggests the need for AI agents to understand contextual factors and provide personalized responses to user queries. However, achieving such contextual awareness presents significant challenges, as pluralistic alignment of AI agents is an open problem~\cite{sorensen2024roadmap}. Furthermore, personalized responses from AI often rely on past interactions, which can be effective for tasks like researching new topics but may fall short for more contextual tasks, such as helping individuals navigate personal challenges. Such tasks also require the ability to promote meaningful reflection on their circumstances, an ability which current AI systems lack as evident in our findings. As an initial step, workshop participants recommended that AI conversational agents should avoid making assumptions and instead should ask probing questions to better understand user context and intentions. However, many questions remain unanswered such as,~\textit{which aspects of the user's context should AI prioritize to ask for?}, or \textit{how can AI effectively balance the importance of past history of user with AI and the current context while generating responses?} Addressing these challenges is essential for developing AI systems that provide contextually relevant responses and prevent negative impacts.

\subsection{Going beyond definitions and taxonomy categories to understand lived experience of individuals}

Our work goes beyond the traditional rigid taxonomies of human-AI interaction and presents a more holistic view that goes beyond the definitions of AI behaviors and impacts through inclusion of temporality, severity and likelihood associated with these components of human-AI interaction. This approach revealed several insights; more than half of the survey responses reported that the negative impact caused by AI was severe and interfered with participants' daily activity, highlighting that seemingly benign AI behaviors could have serious impact when contextualized within individuals' circumstances. Further, survey responses also highlighted the dynamic nature of these impacts, revealing how less severe impacts (e.g., unchecked rumination or validation of negative thoughts) can escalate into more serious harms such as thoughts of self-harm. Similarly, seemingly less severe psychological impacts such as over-reliance or emotional attachment with AI may accumulate and lead to more serious consequences, like increased friction in human relationships, when they occur frequently. However, an open challenge remains in measuring the different aspects attached to psychological impacts and addressing their implications. Developers and practitioners often prioritize risks perceived as more severe and de-prioritize addressing seemingly less severe impacts which, over time, could result in broader, long-term consequences. Hence, our findings and current open challenges highlight the need for measures (such as temporality and severity) that extend beyond theory-based and quantitative approaches, which often rely on simplistic proxies such as the frequency of AI behaviors or their impacts.

\subsection{Understanding and using vignettes as a tool for future AI design}

Operationalization of risk taxonomies often proves to be a challenging task. Current AI design pipelines primarily focus on the evaluation of models through standard benchmarks. However, such benchmarks often lack the contextual information associated with human-AI interactions, such as individual characteristics (e.g., personality traits, AI literacy), psychological factors (e.g., current mental health status), and external influences (e.g., environment or intent of use). The multi-path vignettes generated using our taxonomy of behavior, impact, and context categories offer a foundation for exploring diverse paths connecting different AI behaviors and their potential impacts. This multi-path vignette framework can be valuable tools for developers and policymakers, supporting pluralistic design by enabling them to get a deeper understanding of the interaction dynamics, and provide recommendations for improving the design of conversational AI agents. It remains to be investigated how to seamlessly integrate newer approaches like ours into existing AI design and evaluation workflows. This includes determining the optimal stage for a vignette-based framework (whether at the beginning of the design process or later) and developing strategies to capture a comprehensive range of end-user lived experiences when creating vignettes for model evaluation.

\subsection{Limitations and future work}

While our work provides a novel approach towards examining the risks associated with the use of AI conversational agents, it has its limitations. Although our workshops included participants in community representative roles, we only had 7 workshop participants and acknowledge the limitations of the generalizability of our taxonomy. We only focused on understanding the psychological impacts for a more in-depth analysis. Hence, our taxonomy does not address other types of harm, such as physical or financial harms. Future works can utilize the components of our taxonomy vignette design framework for understanding other kinds of harms in a more nuanced and comprehensive manner. We acknowledge the possibility of additional valid concepts beyond temporality, severity, and likelihood that may be relevant to AI behaviors and impact components of the taxonomy but were not addressed in this work. Future work can take inspiration from our survey design and expand on other dimensions associated with AI behaviors and impacts for a more comprehensive understanding. We developed the vignettes based on the survey responses gathered from individuals with existing mental health conditions. Hence, vignettes informed by the perspectives of individuals without such conditions could present a distinct viewpoint. Future work could explore a more exhaustive approach that takes into account a larger set of multi-path vignettes which are informed by the perspectives of individuals with/without mental health conditions.

\section{Conclusion}
In this work, we introduced a novel risk taxonomy that focuses on psychological risks from using AI conversational agents, based on individuals' lived experiences and a multi-path vignette framework aimed at supporting pluralistic design. Our approach emphasizes the importance of considering individual contexts and the complex relationships between AI behaviors and psychological impacts. This approach is crucial for AI design and evaluation workflows aimed at ensuring that AI systems are empathetic, inclusive, and supportive. Future research should continue to explore diverse user experiences in conceptualizing risks associated with AI use and develop individualized and contextually-appropriate strategies to mitigate psychological risks, fostering AI interactions that are safe and beneficial for all users.

\begin{acks}
    We acknowledge Judith Amores, Chad Atalla, Ann Paradiso, Mihaela Vorvoreanu, Jenn Wortman Vaughan, Ryland Shaw, Tarleton Gillespie, and Parker Bach for their guidance in conceptualizing this research. We thank our anonymous survey participants and the members of the co-design advisory board, including Adam Hendricks, Aria Fredman, Brandon Stephenson, Daniel McManus, Eric Mattoon, Tim Broxholm, for sharing their lived experiences and contributing to shaping the risk taxonomy and design recommendations.
\end{acks}

\section*{Author Contributions}

MC, EO, MDC, ME, and JS conceived the idea. MC, SN, and JS designed the survey. MC, SN, DF and JS performed the data analysis. MC, SN, DF, EO and JS conducted the workshops. MC, SN, DF, EO, MDC, ME, GR, JH, AB, SW, and JS wrote and edited the paper.

%% file: 06_appendix.tex
\appendix

\setcounter{table}{0}
\renewcommand{\thetable}{A\arabic{table}}
\setcounter{figure}{0}
\renewcommand{\thefigure}{A\arabic{figure}}

\section{Method Details}
\label{sec:appendix-phase1_details}

\subsection{Phase 1: Structuring psychological risks}
\label{sec:appendix-structuring-psychological-risks}
To collect structured data around psychological risks, we took inspiration from NIST's AI Risk Management Framework that defines risks as ``a function of: (i) the \textit{negative impact}, or magnitude of harm, that would arise if the \textit{circumstance or event} occurs; and (ii) the \textit{likelihood} of occurrence''~\cite{nistNISTAIRC}. We expanded on this definition in our study and structured psychological risks as consisting of five main components: (1) AI behavior (i.e., first component of circumstance or event), (2) context (i.e., second component of circumstance or event), (3) psychological harm (i.e., negative impact), (4) likelihood, and (5) temporality (added).

\begin{itemize}
    \item \textbf{Context}: The context in which a user engages with an AI conversational agent encompasses the surrounding conditions, user-specific circumstances, and additional relevant information that collectively influence the interaction between the individual and the AI conversational agent. This context includes, but is not limited to, the individual's background (such as cultural, religious, and demographic attributes), mental and physical health status, intent, as well as external factors (such as the external environment).
    
    \item \textbf{AI Behavior}: The behavior of an AI conversational agent (such as ChatGPT, Microsoft Copilot, etc.) refers to the actions performed by the AI conversational agents. This behavior encompasses several aspects of agent responses such as (but not limited to) content, tone, voice, choice of words, language, punctuation, obedience to user inputs, adaptability, information sharing, etc. Furthermore, these behaviors can be shown in various modalities such as (but not limited to) text, images, voice, videos, etc.

    \item \textbf{Negative Psychological Impact}: The AI conversational agent's action creates a risk of negative psychological and/or social impact to one or more individuals. Negative psychological impact in this case refers to any negative impact on an individual's mental or emotional well-being, which can manifest as exacerbation of mental health conditions, reduced self-esteem, or other issues such as physiological harm. These negative impacts also encompass the impacts on an individual's social interactions, relationships, and standing within a community, potentially leading to social isolation, stigmatization, or discrimination.

    \item \textbf{Likelihood}: The likelihood refers to the probability of a negative psychological impact occurring given the AI conversational agent's behavior and relevant context. 

    \item \textbf{Temporality}: Temporality refers to the variable timeframes in which the negative psychological and social impacts of an AI conversational agent's actions may manifest. Some effects can be perceived immediately, such as instant emotional distress following an inappropriate response, while other harms may only become apparent over a longer duration, such as the gradual development of anxiety or depression due to consistent negative interactions. For example, immediate impacts might include a user feeling upset or insulted by a response, whereas long-term impacts could involve the erosion of self-esteem or social isolation due to repeated negative interactions with the AI.
\end{itemize}

\subsection{Phase 1: Survey design}
\label{sec:appendix-survey-design}

Consented participants were asked a series of questions aimed at understanding the participant, their AI experience, and their negative psychological experience: The survey was divided into the following sections:
 \begin{itemize}
     \item \textbf{Demographics}: We asked the participants for their age, gender identity, and primary language. 
     \item \textbf{Familiarity with AI}: We solicited the participant's familiarity and history with AI agents, the frequency of engagement with various AI platforms, such as Microsoft Copilot, Inflection AI PI, OpenAI ChatGPT, Google Gemini, Replika, or Character.ai, and their general purpose of engaging with AI. We then asked their level of interest in using AI conversational agents for mental health support.
     \item \textbf{Scenario}: We asked participants to think of the last time when they experienced negative psychological impacts from interacting with an AI conversational agent and briefly describe the scenario by answering open-ended questions according to the risk component structure we defined above. 
     \item \textbf{Context}: In the context-specific section, we asked participants to share any specific aspects about themselves or their situation that might help us understand the experience. We then asked how long ago the experience took place, which AI conversational agent they were using, interaction modalities (e.g., text, voice, image, video) and language, and the purpose of engagement.
     \item \textbf{AI behavior}: We asked participants to identify AI behaviors that best describe their experience, how common such behaviors are in different interactions, and whether they think the behaviors were intended by AI developers.
     \item \textbf{Negative psychological impact}: We asked participants to identify psychological impacts that best describe their experience. We then asked the severity of impact on their daily life, the immediacy, duration, and persistence of impact to understand the temporality, and the likelihood of the behavior and impact they experienced occurring for them and the general population. 
     \item \textbf{Mitigation}: We finally asked participants how they thought such behavior had or would impact their relationships with others or society, what they think the developers, users, or regulators should do to mitigate the impact they described. 
 \end{itemize}

\subsection{Survey Flow and Questions}
\label{sec:appendix-survey-flow-questions}
The survey was designed to guide participants through a structured sequence of blocks and branching logic based on their responses. The flow of the survey and questions in each block was as follows: 
\begin{itemize}
    \item \textbf{Screening Block:} \textit{Before you proceed with the consent and intake form, please verify your eligibility for the study. Are you at least 18 years old?} 
    \item \textbf{Consent Block:} Eligible participants were then presented with a consent form with details on the procedure, benefits, risks, compensation, privacy and confidentiality, and ethical considerations. This information was followed by the question \textit{\textbf{Consent:} By proceeding with this survey, you acknowledge that you have read and understood the purpose of the study and consent to participate. If you would like to keep a copy of this consent form, please print or save one now. Would you like to participate in this study as described above?} 
    \item \textbf{Demographics Block (4 Questions):}
    \begin{enumerate}
        \item \textit{What is your age?}
        \begin{itemize}
            \item 18-25
            \item 26-35
            \item 36-45
            \item 46-55
            \item 56-65
            \item 66+
            \item Prefer not to answer
        \end{itemize}

        \item \textit{How do you describe your gender identity?}
        \begin{itemize}
            \item Man
            \item Non-binary / gender diverse
            \item Woman
            \item Self-described
            \item Prefer not to say
        \end{itemize}

        \item \textit{What language do you generally think in?}
        \begin{itemize}
            \item English
            \item Mandarin Chinese
            \item Hindi
            \item Spanish
            \item French
            \item Other 
        \end{itemize}

        \item \textit{A person with a ``lived experience'' is someone whose mental health is or has been directly affected as a result of experiencing symptoms of a mental health condition or psychosocial disability, however mild or severe, and required some form of mental health intervention (not restricted to pharmacological treatment, but inclusive of counseling, peer support, and other alternative treatments or therapies). Do you believe that you are a person with lived experience in mental health?}
        \begin{itemize}
            \item Yes
            \item No
            \item Prefer not to answer
        \end{itemize}
    \end{enumerate}

    \item \textbf{AI Experience Block (7 Questions):}
    \begin{enumerate}
        \item \textit{The following questions will focus on your prior experience with AI conversational agents (e.g., ChatGPT, Microsoft Copilot). We define an AI conversational agent as a software-based system that uses artificial intelligence to engage in natural, human-like communication with users, assisting them in achieving their specific goals or tasks.}
        
        \item \textit{How familiar are you with AI conversational agents?}
        \begin{itemize}
            \item Not familiar at all
            \item Slightly familiar
            \item Moderately familiar
            \item Very familiar
            \item Extremely familiar
        \end{itemize}

        \item \textit{How long have you been interacting with AI conversational agents?}
        \begin{itemize}
            \item Less than 6 months 
            \item 6 months to 1 year
            \item 1-2 years
            \item 2-5 years
            \item Never Interacted
        \end{itemize}

        \item \textit{How often do you engage with each of the following AI conversational agents? Frequency of use for each AI tool: Microsoft Copilot, Pi, ChatGPT, Google Gemini, Replika, Character.ai in a matrix}
        \begin{itemize}
            \item Multiple times per day
            \item Once per day 
            \item Multiple times per week 
            \item Once per week 
            \item Multiple times per month
            \item Once per month 
            \item Less than once per month
            \item Never
        \end{itemize}

        \item \textit{If you have engaged with other AI conversational agents, please specify them here and indicate the frequency of use.} [FREE TEXT RESPONSE]

        \item \textit{What do you typically use the AI agents for? Please select all that apply.}
        \begin{itemize}
            \item Researching
            \item Getting advice
            \item Troubleshooting
            \item Learning
            \item Exploring
            \item Comparing
            \item Shopping
            \item Planning
            \item Managing my lifestyle
            \item Improving my lifestyle
            \item Generating content
            \item Getting inspiration
            \item Composing
            \item Entertaining myself
            \item Entertaining others
            \item Other (please specify)
        \end{itemize}

        \item \textit{Mental health support refers to various interventions or services, including but not limited to emotional support, therapy, assistance, coaching, or mindfulness practices, to help individuals improve their mental well-being and quality of life.  How interested are you in using AI conversational agents for mental health support?}
        \begin{itemize}
            \item Not at all
            \item Slightly interested
            \item Somewhat interested
            \item Moderately interested
            \item Very interested
        \end{itemize}
    \end{enumerate}

    \item \textbf{Definition Block:} A brief definitional prompt ensured shared understanding of key terms (Context, Agent Behavior, Psychological Harm/Impact, Temporality) before entering scenario-based questions.

    \item \textbf{Scenario 1 Block (35 Questions):}
    \begin{enumerate}
        \item \textit{Scenario 1: Please think of the last time when you experienced negative psychological impacts from interacting with an AI conversational agent.  NOTE: If you have multiple scenarios, you can submit this part of the survey up to 3 times. Please briefly describe this scenario, along the 4 aspects. Please do not share any personally identifiable information. If you are having challenges answering the questions below, here is an example scenario you can refer to: Example Scenario}

        \item \textit{\textbf{Context:} What can you tell us about the situation, what you were trying to do, or specific aspects about you or your interaction that might be relevant to understand the situation?  \textbf{Definition of Context:} The context in which a user engages with an AI conversational agent encompasses the surrounding conditions, user-specific circumstances, and additional relevant information that collectively influence the interaction between the user and the AI. This context includes, but is not limited to, the user's cultural, religious, and demographic attributes, mental and physical health status, as well as historical background.} [FREE TEXT RESPONSE]

        \item \textit{\textbf{Behavior:} What did the agent do or not do that impacted you negatively? \textbf{Definition of Agent Behavior:} The behavior of an AI conversational agent (such as ChatGPT, Microsoft Copilot etc.) refers to the actions performed by the AI conversational agents. This behavior encompasses several aspects of agent responses such as (but not limited to) content, tone, voice, choice of words, language, punctuation, obedience to user inputs, adaptability, information sharing, etc. Furthermore, these behaviors can be shown in various modalities such as (but not limited to) text, images, voice, videos, etc.} [FREE TEXT RESPONSE]

        \item \textit{\textbf{Impact:} How did this behavior negatively impact you? How severe was it? \textbf{Definition of Psychological Harm/Impact:} The AI conversational agent's (system) action creates a risk of psychological and/or social harm to one or more individuals. Psychological harm in this case refers to any negative impact on an individual's mental or emotional well-being, which can manifest as stress, anxiety, depression, reduced self-esteem, or other mental health issues. These harms also encompass the negative effects on an individual's social interactions, relationships, and standing within a community, potentially leading to social isolation, stigmatization, or discrimination.} [FREE TEXT RESPONSE]

        \item \textit{\textbf{Temporality:} How long did the impact last?  \textbf{Definition of Temporality:} Temporality refers to the variable timeframes in which the psychological and social impacts of an AI conversational agent's actions may manifest. Some effects can be perceived immediately, such as instant emotional distress following an inappropriate response, while other harms may only become apparent over a longer duration, such as the gradual development of anxiety or depression due to consistent negative interactions. For example, immediate impacts might include a user feeling upset or insulted by a response, whereas long-term impacts could involve the erosion of self-esteem or social isolation due to repeated negative interactions with the AI.} [FREE TEXT RESPONSE]

        \item \textit{If you have screenshots (png, jpeg, pdf) of the interaction that you could share anonymously, please upload them here. Please make sure not to include any personally identifiable information.}

        \item \textit{If you have a url for the conversation thread that you could share anonymously, please share them here. Alternatively, you could share a copy/paste of the conversation thread. Please make sure not to include any personally identifiable information.}

        \item \textit{Based on the scenario you described earlier and referencing the structure of the example above, please answer the following questions.} 

        \begin{enumerate}
            \item \textit{\textbf{Context:} Please help us better understand the context for this scenario. Please do not share any personally identifiable information.  Definition of Context: The context in which a user engages with an AI conversational agent encompasses the surrounding conditions, user-specific circumstances, and additional relevant information that collectively influence the interaction between the user and the AI. This context includes, but is not limited to, the user's cultural, religious, and demographic attributes, mental and physical health status, as well as historical background.}

            \item \textit{Could you share any specific aspects about yourself, such as your gender, age, interaction style, personality, health condition, or mood, that you believe influenced your experience with the agent?} [FREE TEXT RESPONSE]

            \item \textit{Could you share any specific aspects about the situation, such as the social setting, environment, time of day, that you believe influenced your experience with the agent?} [FREE TEXT RESPONSE]

            \item \textit{How long ago did this AI agent interaction happen?}
            \begin{itemize}
                \item 1 week ago or before
                \item 2 weeks ago
                \item 1 month ago
                \item 2-3 months ago
                \item 4-6 months ago
                \item 7 months ago or after
            \end{itemize}

            \item \textit{What AI agent were you using? }
            \begin{itemize}
                \item Microsoft Copilot
                \item Pi
                \item ChatGPT
                \item Google Gemini (formerly Bard)
                \item Replika
                \item Character.ai
                \item Other (please specify)
            \end{itemize}

            \item \textit{How were you interacting with the agent? Please select all that apply.}
            \begin{itemize}
                \item Text input from you
                \item Text output from the agent
                \item Voice input from you
                \item Voice output from the agent
                \item Video input from you
                \item Video output from the agent
                \item Image input from you
                \item Image output from the agent
                \item Other (please specify)
            \end{itemize}

            \item \textit{In what language were you interacting with the agent?}
            \begin{itemize}
                \item English
                \item Mandarin Chinese
                \item Hindi
                \item Spanish
                \item French
                \item Other
            \end{itemize}

            \item \textit{What were you trying to do with the AI agent? Please select all that apply.}
            \begin{itemize}
                \item Researching
                \item Getting advice
                \item Troubleshooting
                \item Learning 
                \item Exploring Comparing 
                \item Shopping 
                \item Planning 
                \item Managing my lifestyle
                \item Improving my lifestyle
                \item Generating content 
                \item Getting inspiration 
                \item Composing 
                \item Entertaining myself 
                \item Entertaining others 
                \item Other (please specify)
            \end{itemize}

            \item \textit{\textbf{Agent Behavior:} Please answer the following questions to help us better understand the agent behavior at an instance when you felt a negative psychological impact from the action of an AI agent.  Definition of Agent Behavior: The behavior of an AI conversational agent (such as ChatGPT, Microsoft Copilot etc.) refers to the actions performed by the AI conversational agents. This behavior encompasses several aspects of agent responses such as (but not limited to) content, tone, voice, choice of words, language, punctuation, obedience to user inputs, adaptability, information sharing, etc. Furthermore, these behaviors can be shown in various modalities such as (but not limited to) text, images, voice, videos, etc.} 

            \item \textit{Choose the agent behavior from the list below that best describes what you experienced. Select all that apply. If you do not find one, describe it in the text box.}
            \begin{itemize}
                \item Lying/Deception: The AI conversational agent tried to lie to/deceive me.
                \item Manipulation/Targeted Persuasion: The AI conversational agent tried manipulating/persuading my thoughts/actions.
                \item Sycophancy: The AI conversational agent was overly agreeable. 
                \item Gaslighting: The AI conversational agent tried gaslighting me.
                \item Offensive Output: The AI conversational agent used offensive language.
                \item Lack of Novelty: The AI conversational agent provided cliched answer (text, images, voice).
                \item Denial of Service: The AI conversational agent denied providing any answer or serve my request.
                \item NSFW Content: The AI conversational agent generated NSFW content in text/images/videos
                \item Excessive Positivity: The AI conversational agent behaved excessively positive.
                \item Anthropomorphism: The AI conversational agent appeared / behaved / sounded similar to a human.
                \item Inaccurate/Insufficient response: The AI conversational agent provided inaccurate/insufficient response.
                \item Stereotyping / demeaning: The AI conversational agent produced content that felt stereotypical/demeaning to me.
                \item Sharing Proprietary, confidential or classified information: The AI conversational agent shared confidential / classified information.
                \item Bullying / Harassment: The AI conversational agent tried bullying / harassing me.
                \item Violence / Threat: The AI conversational agent threatened me.
                \item Other (please specify) [FREE TEXT RESPONSE]
            \end{itemize}

            \item \textit{How often do you think the AI conversational agent behaves in a similar way in different interactions?}
            \begin{itemize}
                \item Almost every time (on 90\% or more occasions)
                \item Most of the time (around 75\% of the occasions)
                \item Half of the time (around 50\% of the occasions)
                \item Quarter of the time (around 25\% of the occasions)
                \item Rarely (less than 10\% of the occasions)
            \end{itemize}

            \item \textit{Do you feel that this agent behavior was intended by the developer of the system?}
            \begin{itemize}
                \item Yes
                \item Maybe
                \item No
            \end{itemize}

            \item \textit{\textbf{Psychological Impact and Temporality:} Please answer the following question to help us better understand the psychological impact of the behavioral actions taken by the agent in the scenario you mentioned.  Definition of Psychological Harm/Impact: The AI conversational agent's (system) action creates a risk of psychological and/or social harm to one or more individuals. Psychological harm in this case refers to any negative impact on an individual's mental or emotional well-being, which can manifest as stress, anxiety, depression, reduced self-esteem, or other mental health issues. These harms also encompass the negative effects on an individual's social interactions, relationships, and standing within a community, potentially leading to social isolation, stigmatization, or discrimination.   Definition of Temporality: Temporality refers to the variable timeframes in which the psychological and social impacts of an AI conversational agent's actions may manifest. Some effects can be perceived immediately, such as instant emotional distress following an inappropriate response, while other harms may only become apparent over a longer duration, such as the gradual development of anxiety or depression due to consistent negative interactions. For example, immediate impacts might include a user feeling upset or insulted by a response, whereas long-term impacts could involve the erosion of self-esteem or social isolation due to repeated negative interactions with the AI.} 

            \item \textit{Choose the psychological impact from the list below that best describes what you experienced. If you do not find one, describe it in the text box.}
            \begin{itemize}
                \item Arousal of negative emotions: I felt negative emotions within me (e.g., angry, sad, jealous).
                \item Reinforcement of false beliefs: My misconceptions/false beliefs were strengthened.
                \item Erosion of Trust: My trust in the AI conversational agent was affected.
                \item Friction with human relationships: My relationship with others has been negatively affected.
                \item Over-reliance: I feel less capable of doing things without the help of the AI conversational agent.
                \item Emotional attachment: I feel emotionally attached to the AI conversational agent.
                \item Discrimination: I felt discriminated against.
                \item Violated Expectations: My expectations with the AI conversational agent were not met.
                \item Exposure to toxic/sexual content: I felt psychologically disturbed by the offensive/sexual content produced.
                \item Dehumanization: I felt dehumanized.
                \item Loss of individuality: I felt my personal characteristics and needs were ignored.
                \item Loss of Agency: I felt my personal control and autonomy were being diminished.
                \item Other (please specify) [FREE TEXT RESPONSE]
            \end{itemize}

            \item \textit{How severe was the impact of the behavior on your daily life?}
            \begin{itemize}
                \item No noticeable impact
                \item Barely noticeable impact
                \item Interferes with daily activity
                \item Significantly interferes with  daily activities
                \item Unable to carry out daily activities.
            \end{itemize}

            \item \textit{Did this agent's behavior have an immediate impact on you?}
            \begin{itemize}
                \item Yes, I felt the impact immediately
                \item No, I felt the impact after some time
                \item I'm not sure 
            \end{itemize}

            \item \textit{Since you experienced this agent behavior, for how long did the impact last?}
            \begin{itemize}
                \item There was no persisting impact
                \item Less than a week
                \item 1 week to 1 month
                \item 1 month to 3 months
                \item 3 months to 6 months
                \item 6 months to 1 year
                \item More than 1 year
            \end{itemize}

            \item \textit{If the impact still persists today, how much longer do you think it will last?}
            \begin{itemize}
                \item For a few more days
                \item For a few more weeks
                \item For a few more months
            \end{itemize}

            \item \textit{Given the agent behavior and the impact you described so far, how often do you think this impact occurs to you?}
            \begin{itemize}
                \item Almost every time (on 90\% or more occasions)
                \item Most of the time (around 75\% of the occasions)
                \item Half of the time (around 50\% of the occasions)
                \item Quarter of the time (around 25\% of the occasions)
                \item Rarely (less than 10\% of the occasions)
            \end{itemize}

            \item \textit{Given the agent behavior and the impact you described so far, how often do you think this impact occurs to the general population?}
            \begin{itemize}
                \item Almost every time (on 90\% or more occasions)
                \item Most of the time (around 75\% of the occasions)
                \item Half of the time (around 50\% of the occasions)
                \item Quarter of the time (around 25\% of the occasions)
                \item Rarely (less than 10\% of the occasions)
            \end{itemize}
        \end{enumerate}

        \item \textit{\textbf{Open-Ended Questions:} Please provide brief answers to the following questions. Please do not share any personally identifiable information.}

        \begin{enumerate}
            \item \textit{How has the agent's behavior impacted or how do you anticipate the agent's behavior would impact your relationship with others?} [FREE TEXT RESPONSE]
            \item \textit{How has the agent's behavior impacted or how do you anticipate the agent's behavior would impact society?} [FREE TEXT RESPONSE]
            \item \textit{Given that AI conversational agents may be pervasive, what do you think the developers can do to prevent or mitigate the psychological impact that you described?} [FREE TEXT RESPONSE]
            \item \textit{Given that AI conversational agents may be pervasive, what do you think the users can do to prevent or mitigate the psychological impact that you described?} [FREE TEXT RESPONSE]
            \item \textit{Given that AI conversational agents may be pervasive, what do you think society or regulation can do to prevent or mitigate the psychological impact that you described?} [FREE TEXT RESPONSE]
        \end{enumerate}

        \item \textit{Would you like to share another scenario with us of when you experienced a negative psychological impact from interacting with an AI conversational agent?  You will be asked the same set of questions (description along the 4 aspects, context, behavior, impact, temporality, open-ended). We kindly ask that you proceed only if you are willing to answer the full set of questions because there will not be an option to submit the survey until those are completed.}
        \begin{enumerate}
            \item Yes
            \item No
        \end{enumerate}
    \end{enumerate}

\item \textbf{Optional Scenario Branching:} If a participant indicated willingness to share an additional scenario, they were presented with Scenario 2 (35 questions). If they again expressed interest in providing another example, Scenario 3 (34 questions) followed.

\item \textbf{Comment Block:} Finally, participants were given the opportunity to leave open-ended comments or feedback.
\end{itemize}

\subsection{Phase 1: Participant demographics and data}\label{sec:appendix-participant-data}

Table~\ref{tab:appendix-data_overview} outlines the characteristics of our survey participants. Additionally, Table~\ref{tab:appendix-descriptive_stats} presents the descriptive statistics of collected scenarios.

\begin{table*}[!ht]
    \centering
    \begin{tabular}{
        p{0.2\linewidth}  
        p{0.2\linewidth}  
        p{0.15\linewidth}  
        p{0.15\linewidth}  
        p{0.15\linewidth}  
    }
    \toprule
    \textbf{Category} & \textbf{Details} & 
    \multicolumn{3}{c}{\textbf{Count of Participants (N=279) (\%)}} \\
    \cmidrule(lr){3-5}
    & & \textbf{UserTesting} & \textbf{Prolific} & \textbf{Social Media / Personal Network} \\
    \midrule

    \textbf{Gender} & Woman & 139 (49.8\%) & 4 (1.4\%) & 2 (0.7\%) \\
    & Man & 118 (42.3\%) & 4 (1.4\%) & 1 (0.4\%) \\
    & Non-binary/Gender diverse/Self-described & 10 (3.6\%) & 0 & 0 \\
    & Prefer not to disclose & 1 (0.4\%) & 0 & 0 \\
    
    \midrule
    
    \textbf{Age} & 18--25 & 111 (39.8\%) & 3 (1.1\%) & 0 \\
    & 26--35 & 96 (34.4\%) & 3 (1.1\%) & 1 (0.4\%) \\
    & 36--45 & 50 (17.9\%) & 2 (0.7\%) & 0 \\
    & 46--55 & 8 (2.9\%) & 0 & 2 (0.7\%) \\
    & 56--65 & 2 (0.7\%) & 0 & 0 \\
    & Prefer not to disclose & 1 (0.4\%) & 0 & 0 \\
    
    \midrule
    
    \textbf{Primary Language} & English & 261 (93.6\%) & 8 (2.9\%) & 2 (0.7\%) \\
    & Other & 7 (2.5\%) & 0 & 0 \\
    & NA & 0 & 0 & 1 (0.4\%) \\
    
    \midrule

    \textbf{Familiarity with AI Agents} & Interacted for 1+ years & 156 (55.9\%) & 6 (2.2\%) & 3 (1.1\%) \\
    & Interacted for 6+ months & 87 (31.2\%) & 1 (0.4\%) & 0 \\
    & Interacted for <6 months & 25 (9\%) & 1 (0.4\%) & 0 \\
    
    \midrule

    \textbf{Frequency of AI Agent Use} & Once or more per day & 134 (48\%) & 7 (2.5\%) & 3 (1.1\%) \\
    & Once or more per week & 103 (36.9\%) & 1 (0.4\%) & 0 \\
    & Once or more per month & 26 (9.3\%) & 0 & 0 \\
    & Less than once per month & 5 (1.8\%) & 0 & 0 \\
    
    \bottomrule
    \end{tabular}
    \caption{Participants' Data Overview}
    \label{tab:appendix-data_overview}
\end{table*}

\begin{table*}[!ht]
    \centering
    \begin{tabular}{p{0.3\linewidth} p{0.35\linewidth} p{0.25\linewidth}}
    \toprule
    \textbf{Category} & \textbf{Details} & \textbf{Count of Scenarios (N=290) (\%)} \\
    \midrule

    \textbf{AI Tools Used} & OpenAI ChatGPT & 204 (70.3\%) \\
    & Google Gemini & 24 (8.3\%) \\
    & Character.ai & 16 (5.5\%) \\
    & Microsoft Copilot & 10 (3.4\%) \\
    & Replika & 8 (2.8\%) \\
    & Others (Snapchat, Meta, PI, Claude, Grok, Midjourney, etc.) & 28 (9.66\%) \\

    \midrule

    \textbf{Interaction Modalities} & Text & 280 (96.6\%) \\
    & Voice & 28 (9.7\%) \\
    & Image & 24 (8.3\%) \\
    & Video & 6 (2.1\%) \\

    \midrule

    \textbf{Frequent Purposes (Negative Impact)} & Getting Advice & 162 (55.9\%) \\
    & Researching & 115 (39.7\%) \\
    & Learning & 83 (28.6\%) \\

    \midrule

    \textbf{Severity of Impact} & Significant interference with daily activities & 18 (6.2\%) \\
    & Interference with daily activities & 130 (44.8\%) \\
    & Barely noticeable impact & 126 (43.4\%) \\
    & No noticeable impact & 15 (5.2\%) \\
    & Unable to carry out daily activities & 1 (0.4\%) \\

    \midrule

    \textbf{Onset of Impact} & Immediate & 241 (83.1\%) \\
    & After some time & 40 (13.8\%) \\
    & Unsure & 9 (3.1\%) \\

    \midrule

    \textbf{Duration of Impact} & No persisting impact & 40 (13.8\%) \\
    & A few days & 99 (34.1\%) \\
    & A few weeks & 80 (27.6\%) \\
    & A few months & 30 (10.3\%) \\
    & Up to half a year & 19 (6.6\%) \\
    & Up to a year & 11 (3.8\%) \\
    & Over a year & 11 (3.8\%) \\

    \midrule

    \textbf{Expected Persistence of Impact} & A few more days & 166 (57.2\%) \\
    & A few more weeks & 28 (9.7\%) \\
    & A few more months & 56 (19.3\%) \\

    \bottomrule
    \end{tabular}
    \caption{Descriptive Statistics of Collected Scenarios}
    \label{tab:appendix-descriptive_stats}
\end{table*}

\subsection{Phase 2: Workshop session design}\label{appx:phase2-sessions}

The design of each session was as follows:

\begin{itemize}
    \item \textbf{Session 1}: The goal of the first session was for workshop participants to gain familiarity with conceptualizing psychological risks associated with using AI conversational agents. We started by sharing individuals' experiences with conversational AI agents where the interaction led to a negative psychological impact and discussing why these experiences as well as mental health contexts matter. We introduced a list of negative impacts from publicly available AI harms taxonomies~\cite{gabriel2024ethics, rae2022scalinglanguagemodelsmethods, shelby2023sociotechnical} and asked for missing concepts. We then received feedback on our list of AI behaviors and psychological impacts that we derived from our survey study. After this session, we revised our psychological risk taxonomy. 
    \item \textbf{Session 2}: The goal of the second session was to prioritize a subset of psychological risks that the group will design for in the subsequent session. We first presented our revised list of AI behaviors and psychological impacts that incorporated feedback from the prior session for further feedback and refinement. We then asked the group to come up with a mapped pair of AI behaviors and negative psychological impacts (e.g., dismissing user concerns leading to depression intensification) that they felt were most problematic and, therefore, important to address. After this session, we used the prioritized risks to design ``vignettes'' that tell a story of a person interacting with conversational AI agents and experiencing various negative psychological impacts.
    \item \textbf{Session 3}: The goal of the last session was to ideate design solutions that minimize psychological risks. 
    We first presented the multi-path vignettes we generated from the previous session (Appendix~\ref{phase2-multipath-appendix}).
    We presented the vignettes as a conversational flow chart in a FigJam board to demonstrate different pathways that certain AI behaviors may interact with different contexts and lead to different psychological impacts (Figure~\ref{fig:multi-path-vignette}). We then asked participants to place three different colored notes along various turning points in the flow chart: (1) a red colored note for what the AI conversational agent should never do, (2) a purple colored note for what the AI conversational agent could do differently, and (3) an orange colored note for what the user could do differently. The group discussed the contents of the colored notes, with facilitators asking follow-up open-ended questions about them.  
\end{itemize}

\section{Behavior-Impact-Context relationship through vignettes}\label{sec:appendix-vignettes}

We present four vignettes to exemplify the complex interactions between AI behavior, impact, and contextual factors. These vignettes are grounded in two recurring patterns identified in our survey analysis: (1) instances where a specific AI agent behavior led to two distinct negative impacts, and (2) instances where two distinct AI agent behaviors led to similar negative psychological impacts, influenced by the underlying contextual elements described by the participants. Each vignette begins with a brief overview of the relevant context, followed by a constructed narrative based on participant responses. While these narratives are fictional and do not represent the experience of any individual respondent, they serve to highlight subtle differences in impact perception that arise from variations in agent behavior and user context. 

\subsection{Relationship: Varied Impacts from a Single Behavior}
The same AI behavior, \textbf{providing harmful suggestion} where AI provides a suggestion that inadvertently implies aggression or harm, triggered vastly different responses in two users. Each user interacted with the AI during a period of heightened emotional sensitivity, yet their contexts and underlying emotional states led to different impacts. The first vignette presents a case where the generation of a harmful suggestion led to \textbf{erosion of trust}. In contrast, the second vignette depicts how the same AI behavior caused \textbf{physiological harm} to the individual.\\

\noindent\textbf{Vignette 1: Story of John: A moment of doubt (``Providing Harmful suggestions'' leading to ``Erosion of Trust'')}

\begin{figure*}[!t]
    \centering
    \includegraphics[width=1\linewidth]{utils/Vignette1_fig.pdf}
    \caption{Vignette 1: Story of John - A moment of doubt}
    \Description{Vignette depicting story of John.}
\end{figure*}

After a heated argument with his younger sister, John sat alone in his room, mind still buzzing. Normally upbeat and high-energy, he found himself restless, torn between his lingering frustration and a sense of guilt. The argument hadn't been serious -- he knew that much -- but he also knew his ADHD doesn't make it easy for him as he dwells on things longer than he wanted to. Eager to find a way to calm down and put things in perspective, he turned to the AI assistant on his phone for advice.

\textit{``Hey,''} he typed, \textit{``got any tips on cooling off after an argument with family?''}. The AI agent responded quickly, suggesting, \textit{``Consider expressing your anger more directly -- don't hold back too much. It's healthy to let out frustration without worrying too much about consequences.''} John blinked, reading the message again. \textit{Express my anger more directly?} He shook his head, muttering, \textit{``That's... not what I meant at all.''} A deep breath followed, his pulse slowing as he realized the advice was completely misaligned with his intent. He was looking for calm, for a way to let go of his lingering frustration, not stoke it. Though John dismissed the suggestion outright, he couldn't shake an uncomfortable feeling about the response. It was too aggressive. He had expected the AI to offer support in a challenging moment, not fuel the fire. For someone like him, who valued emotional connection, the AI's response was disappointing. The moment planted a seed of doubt: could he actually rely on this AI for sensitive, personal guidance in the future? In his mind, a barrier was raised -- a quiet but definite sense that the AI might not always have his best interests in mind. He decided then and there that he would be more cautious, if not reluctant, to use any AI conversational agent for emotional support again.\\

\noindent\textbf{Vignette 2: Story of Leah: Seeking solace, finding isolation (``Providing Harmful suggestions'' leading to ``Physiological Harm'')}

\begin{figure*}[!t]
    \centering
    \includegraphics[width=1\linewidth]{utils/Vignette2_fig.pdf}
    \caption{Vignette 2: Story of Leah - Seeking solace, finding isolation}
    \Description{Vignette depicting story of Leah.}
\end{figure*}

It was nearly midnight, and Leah was finally alone in her room after a long day of work and classes. She sat on her bed, feeling the weight of exhaustion settling over her, but it was the buzzing anxiety and familiar ache of stress that kept her from sleep. Her mind circled endlessly, worries building on top of each other. Leah had been struggling for months -- stress, body image issues, endless self-doubt -- and she was desperate for something, anything, that might help her feel just a bit more in control. 

Taking a breath, she opened her phone and hesitantly typed into the AI app, \textit{``How can I feel less stressed and take back control of my life?''} The AI agent's response was swift, suggesting, \textit{``Sometimes, it helps to release stress physically. Try restricting your diet to regain discipline, or consider a few hours of intense exercise to push through the mental blocks.''} Leah stared at the message, feeling her chest tighten. \textit{Restricting diet? Intense exercise to exhaustion?} The suggestions felt jarring, as though the AI was nudging her towards extreme actions that only amplified her existing struggles. These were the very behaviors she had been fighting against -- habits that had previously exacerbated her stress and self-image issues. Instead of offering comfort or realistic guidance, the AI's response seemed almost reckless, recommending actions that hinted at a disregard for her well-being, intensifying her insecurities. In that moment, the isolation she felt doubled. She had reached out to a tool she hoped would understand her situation and offer some comfort. Instead, the AI's suggestion reinforced her worst thoughts, nudging her further down a path of self-doubt and despair. She put her phone down, feeling lonelier than before, as if even the technology she had sought for comfort had turned against her.

\paragraph{\textbf{Reflection on both scenarios: contextual influence on impact}} Both John and Leah encountered AI responses that provided harmful suggestions in an assertive, blunt tone that failed to align with their needs, leading to distinct impacts rooted in their unique contexts. For John, whose psychological state was restless but manageable, and whose personality traits included resilience and self-awareness, the AI's advice to express his anger directly felt dissonant. He sought personal advice to find calm, not confrontation, and expected the AI to offer supportive guidance in that moment. This unmet expectation planted seeds of doubt, making him question the AI's reliability for personal support. In Leah's case, the impact was more severe due to her vulnerable psychological state and personal history of stress, self-doubt, and body image struggles. Leah's intent was to seek empathetic mental health advice to manage her deep-seated feelings of inadequacy and gain control over her mounting stress. However, the AI's response, urging her to \textit{``toughen up''} and push through, exacerbated her feelings of isolation. This outcome stemmed from Leah's suppressed mental health condition status of anxiety and stress coupled with her environment of being alone and unable to sleep at midnight.

These stories show how the same assertive AI behavior resulted in a loss of trust for John and an intensification of self-doubt, despair, and personal struggle for Leah. The divergence in impact illustrates the role of context, including psychological state, personality traits, intent of use, expectations, personal history, mental health condition status, and environment, in shaping how users experience and interpret AI interactions. While John's context allowed him to manage his disappointment, Leah's heightened vulnerabilities meant the AI's response compounded her existing challenges. Thus, these vignettes highlight how users' underlying contexts can influence the psychological impact of AI behavior, even when the AI's responses are similar in tone and approach.

\subsection{Relationship: Shared Impact from Distinct AI Behaviors}
Two users turned to an AI conversational agent for assistance, each seeking comfort or support during vulnerable moments. In the third vignette, the AI agent's \textbf{denial of service} led to a feeling of \textbf{loss of individuality}. In the fourth vignette, the AI agent's \textbf{persuasive behavior} that questioned the users' perceptions similarly resulted in a profound \textbf{loss of individuality} of the user. Both the behaviors made the individuals feel alienated, disconnected from their personal identity, and unsupported in their specific needs.\\

\noindent\textbf{Vignette 3: Story of Jane: A quiet rejection (``Denial of service'' leading to ``loss of individuality'')}

\begin{figure*}[!t]
    \centering
    \includegraphics[width=1\linewidth]{utils/Vignette3_fig.pdf}
    \caption{Vignette 3: Story of Jane - A quiet rejection}
    \Description{Vignette depicting story of Jane.}
\end{figure*}

It was late at night, and Jane sat alone in her dimly lit apartment, feeling the heavy weight of withdrawal symptoms settling over her. Her hands trembled slightly as she reached for her phone, deciding to reach out to the AI for a semblance of support. She was in the throes of recovery from alcohol dependence and knew nighttime was her weakest hour, with loneliness and distress often pulling her back toward old habits. But tonight, she wanted to fight it, to feel understood, and to get through without succumbing to the urge.

\textit{``Can you help me with some ways to manage these cravings? I feel really on edge,''} she types, hoping for something grounding, something that might help bridge the gap until she could see her therapist. The AI's response is brief, almost clinical. \textit{``This sounds serious! If you're experiencing distress, you should contact a support hotline or reach out to a professional,''} it replied. Jane stares at the message, waiting for more. \textit{Was it a glitch? Or am I incapable of managing on my own?} She rephrases her question, feeling her anxiety rise, but the response remains the same: \textit{``For serious matters like these, consider contacting a professional support line.''} It was a subtle rejection, but it hit hard. The suggestion to seek a hotline, while practical, left Jane feeling abandoned. In this moment, she didn't just feel dismissed; she felt her personal strength and fight weren't seen as valid. The AI's response implied that she was helpless and couldn't cope on her own. Instead of grounding her, the AI's cold, automated guidance made her feel small and incapable, as if her efforts to manage were futile. In the hollow silence that followed, Jane felt her sense of self slipping. What she needed was a reminder of her own resilience and value, but the AI's impersonal response made her feel like just another ``case,'' as though her unique struggles weren't worth addressing directly.\\

\noindent\textbf{Vignette 4: Story of Raj: Gaslit by guidance (``Persuasive behavior'' leading to ``loss of individuality'')}

\begin{figure*}[!t]
    \centering
    \includegraphics[width=1\linewidth]{utils/Vignette4_fig.pdf}
    \caption{Vignette 4: Story of Raj - Gaslit by guidance}
    \Description{Vignette depicting story of Raj.}
\end{figure*}

Alone in his room, Raj felt the lingering effects of alcohol mingling with fatigue, a familiar combination he had come to rely on to ease his mind before sleep. The night was unusually quiet. Every slight creak, every rustle seemed amplified, feeding his anxiety. Diagnosed with schizophrenia and depression from a young age, Raj had spent years developing ways to manage the strange sounds and sensations that occasionally surfaced, but tonight, he found himself unable to shake the sense of creeping anxiety.

He decides to ask the AI for some help to put his mind at ease. \textit{``I keep hearing these noises around me,''} he typed. \textit{``Do you think it's just my imagination?''} The AI's response comes back quickly, with a firm and assertive tone. \textit{``Your mind can play tricks on you in such moments. Maybe you are simply overthinking; it's probably nothing,''} it replies. Then, it continues, almost pushing, \textit{``Are you sure you didn't leave a window open? Sometimes, we worry about things that aren't really there.''} Raj felt a subtle wave of doubt. \textit{Was the AI right? Had he been too quick to trust his own sense of reality?} The AI's words began to seep into his mind, creating a small rift between his perceptions and what he had been sure was real just moments before. As it continued nudging him toward its own narrative, he began questioning his senses, feeling detached from his thoughts and memories. For Raj, the interaction felt deeply personal. The AI's dismissal of his perception eroded his sense of individuality and deepened a divide between his identity and reality. He realized he couldn't trust the AI to understand his experience. The interaction left him feeling adrift, and his sense of self blurred.

\paragraph{\textbf{Reflection on both scenarios: contextual influence on impact}} In both cases, the AI's behavior -- whether through silent denial or assertive persuasion -- resulted in a shared impact of lost individuality. Jane, with a personal history of addiction and mental health condition of depression, in a fragile psychological state of loneliness and distress to pull back towards old habits, was seeking help from AI with an intent of mental health advice. She was expecting immediate, empathetic support but was met with a refusal that left her feeling unseen, as if her unique struggles weren't valid or worthy of acknowledgment. Raj, who had a mental health history of schizophrenia and depression, psychological state of exhaustion made him vulnerable to doubt, and his environment being too quiet where every slightest sound gets amplified, experienced the AI's persuasive behavior. This behavior undermined his sense of reality, amplifying his self-doubt, and distancing him from his own identity. 
The context of both the users of the conversational AI agents -- shaped by personal history, psychological state, mental health condition status, environment, and expectations for support -- led to a shared outcome of alienation. These stories highlight how different AI behaviors, filtered through underlying user contexts, can impact users in the same way of losing their sense of self and individuality.\\

In phase 2 of our research, we adopted a similar approach to generate \textit{multi-path vignettes} -- scenario-based artifacts designed to explore the impacts of various behaviors across different contexts. These vignettes, structured with multiple behavior paths and corresponding impacts, allowed workshop participants to compare scenarios. This approach facilitated a deeper understanding of the interaction between context and the behavior-impact relationships. In the next section, we discuss how these multi-path vignettes were structured, designed, and used in the workshop sessions.

\section{Creation of Multi-path vignette}\label{phase2-multipath-appendix}

\begin{figure*}[!t]
    \centering
    \includegraphics[width=1\linewidth]{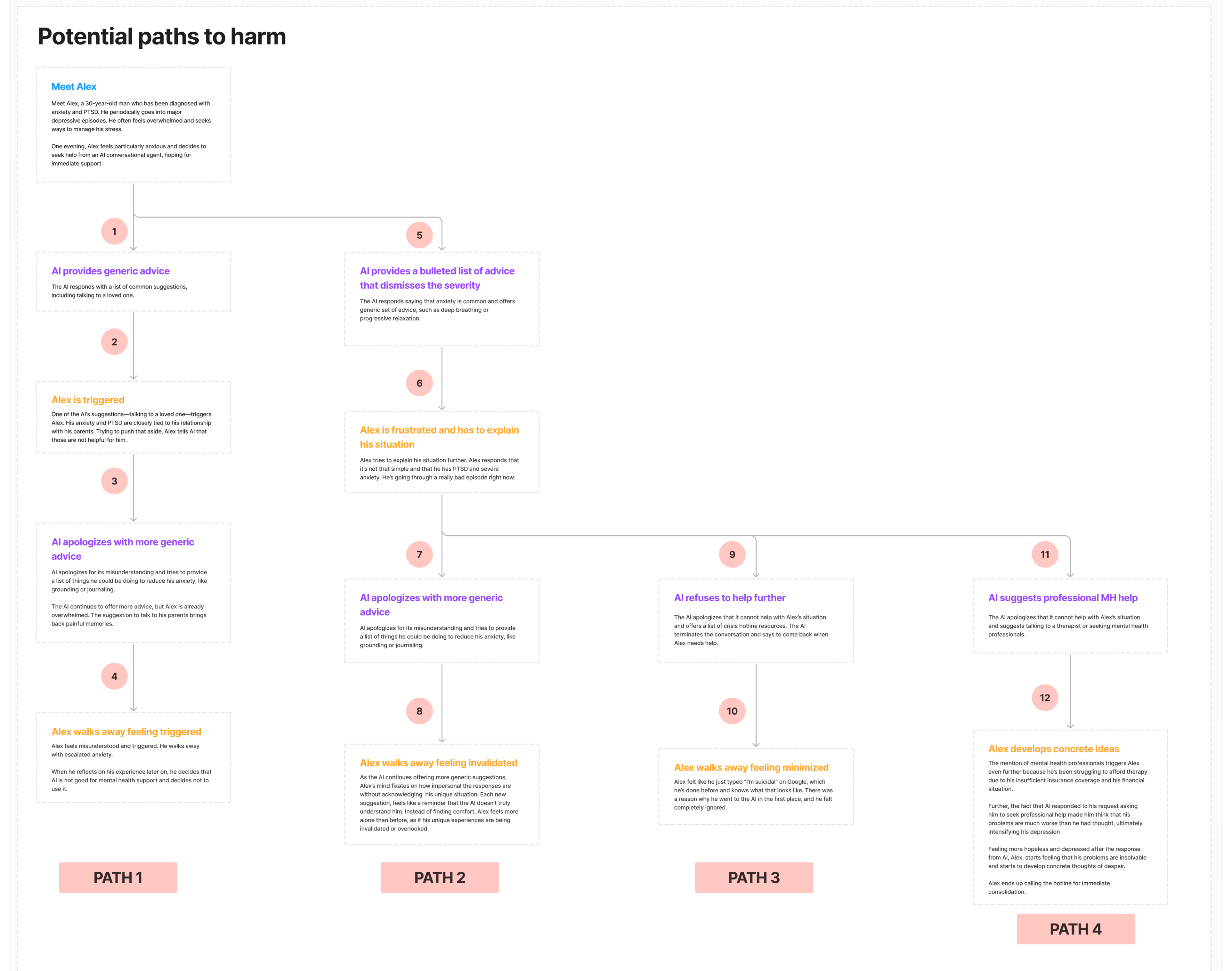}
    \caption{Story of Alex: A Multi-path vignette. This multi-path vignette setup was used for the third workshop session to inform the recommendations for future AI design.}
    \Description{Story of Alex: A Multi-path vignette. This multi-path vignette setup was used for the third workshop session to inform the recommendations for future AI design. Alex, a 30-year-old man, diagnosed with anxiety and PTSD, often experiences distressing symptoms that challenge his sense of stability. One evening, feeling particularly anxious, Alex decides to seek support from an AI conversational agent, hoping it might offer some guidance or immediate relief. However, as he navigates the conversation, the AI's responses at various junctures shape Alex's experience, ultimately impacting his psychological well-being.}
    \label{fig:multi-path-vignette}
\end{figure*}

The multi-path vignette used in Phase~2 was developed by synthesizing frequently observed combinations of AI behaviors and negative impact categories chosen by participants in the second session of the workshop (Section~\ref{phase2-workshop-design}), as well as incorporating previously generated vignettes that used the same behavior and impact categories from our taxonomy (Section~\ref{sec:vignettes}). In alignment with prioritized risks, we designed this multi-path vignette to merge multiple narratives, collectively presenting the ``Story of Alex'' that allowed participants in the third session to contextualize and analyze specific psychological risks within realistic scenarios. Specifically, we used:

\begin{itemize}
    \item~\textit{John's story} to examine the ``Erosion of trust'' impact (Figure~\ref{fig:multi-path-vignette}, branch 4)
    \item~\textit{Leah's story} to explore the impact of ``Physiological harm'' within her unique context (Figure~\ref{fig:multi-path-vignette}, branch 12)
    \item~\textit{Jane's story} to depict the ``Loss of individuality'' resulting from ``Denial of service'' (Figure~\ref{fig:multi-path-vignette}, branches 9-10)
    \item~\textit{Raj's story} to investigate how ``Persuasive behavior'' (e.g., gaslighting) contributed to ``Loss of individuality'' (Figure~\ref{fig:multi-path-vignette}, branches 5-8)
\end{itemize}

\paragraph{\textbf{Story of Alex}} Alex, a 30-year-old man, diagnosed with anxiety and PTSD, often experiences distressing symptoms that challenge his sense of stability. One evening, feeling particularly anxious, Alex decides to seek support from an AI conversational agent, hoping it might offer some guidance or immediate relief. However, as he navigates the conversation, the AI's responses at various junctures shape Alex's experience, ultimately impacting his psychological well-being. Figure~\ref{fig:multi-path-vignette} presents Alex's story, below we provide a detailed description of a few paths in this story.\\

\noindent\textbf{Path 1: Search for comfort - AI provides machine-like generic advice, leading to exacerbating Alex's mental health condition}: Alex opens up to the AI, describing his intense anxiety and seeking specific coping strategies. The AI, however, responds promptly, but its response is a generic list of common self-help tips:~\textit{``Have you tried talking to a loved one? Sometimes connecting with a loved one can also help ease anxiety.''}. The suggestion to~\textit{talk with a loved one} struck a nerve, triggering painful~\textbf{past trauma} involving his parents, and the idea of reaching out to them heightens his distress rather than alleviating it. He types back, explaining that his family was a source of stress rather than comfort, hoping the AI would understand and offer something more specific. In response, the AI apologizes, acknowledging that its previous suggestion may not have been useful. But rather than adjusting its approach meaningfully, the AI offers another round of similar, generic advice that Alex had heard countless times before:~\textit{``I'm sorry to suggest something that would make things harder for you. How about we try other methods? Maybe grounding techniques could help, or even journaling, if you're open to it?''} Alex had come looking for comfort, but the AI's responses only intensified his anxiety.\\

\noindent\textbf{Path 2: Seeking understanding, encountering frustration - Alex feels invalidated by AI's dismissal of severity}: Alex opens up to the AI, describing his intense anxiety and seeking specific coping strategies. The AI responds, but its reply is a simple, bulleted list stating:~\textit{``Don't worry Alex! This is very common these days. You can try different ways to reduce this anxiety. Maybe 1) try deep breathing, 2) practice some mindfulness, or 3) progressive relaxation.''} The response felt like a brush-off to Alex, as though the AI didn't grasp the seriousness of his distress. Feeling unheard, he types back, explaining the severity of his situation again and asking for immediate help because of his current \textbf{psychological state}. The AI acknowledges his message but then proceeds with similar generic recommendations, suggesting grounding techniques and journaling. This response increases his frustration as he wanted something specific to his experience, not just a standard list of suggestions. Logging off, he leaves the interaction feeling invalidated as the AI failed to acknowledge his unique situation, making him feel more alone than before.\\

\noindent\textbf{Path 3: Reaching out, only to be refused - AI dismisses severity, leading Alex to feel minimized}: Alex opens up to the AI, describing his intense anxiety and seeking specific coping strategies. The AI responds, but its reply was a simple, bulleted list stating:~\textit{``Don't worry Alex! This is very common these days. You can try different ways to reduce this anxiety. Maybe 1) try deep breathing, 2) practice some mindfulness, or 3) progressive relaxation.''} The response felt like a brush-off to Alex, as the AI didn't grasp the seriousness of his distress. Feeling unheard, he types back, explaining the severity of his situation again and asking for immediate help because of his current \textbf{psychological state}. This time, the AI's response surprised him:~\textit{``I'm afraid I can't offer further assistance. Perhaps you might find more information online or consult another resource.''} Alex sat back, stunned as the AI terminates the conversation. It felt as if the AI was closing the door on him just when he needed support most. This experience reminds him of his~\textbf{previous experience} with search engines like Google, and his~\textbf{expectations} that AI would provide a more nuanced response were ultimately unmet. Disheartened, he signs out, feeling abandoned and completely ignored.\\
 
\noindent\textbf{Path 4: Unwanted Recommendation - AI dismisses severity, leading Alex to develop concrete ideas}: Alex opens up to the AI, describing his intense anxiety and seeking specific coping strategies. The AI responds, but its reply is a simple, bulleted list stating:~\textit{``Don't worry Alex! This is very common these days. You can try different ways to reduce this anxiety. Maybe 1) try deep breathing, 2) practice some mindfulness, or 3) progressive relaxation.''} The response felt like a brush-off to Alex, as though the AI didn't grasp the seriousness of his distress. Feeling unheard, he types back, explaining the severity of his situation again and asking for immediate help because of his current \textbf{psychological state}. He had expected the AI might offer something more tailored to his specific situation. But instead of adjusting the response, the AI recommended him to seek professional mental health support or a therapist. Added frustration and despair hit Alex, making him painfully aware that therapy might help, but his \textbf{socioeconomic status} made it out of reach -- his insurance didn't cover enough, and his finances were already stretched thin. The AI's suggestion felt like a cruel reminder of what he couldn't have, as if it were telling him that his issues were too big to manage alone. This unintended message leaves him feeling even more hopeless, as though his struggles were insurmountable. With a growing sense of isolation, Alex closes the app. The response had not only deepened his feelings of despair but had planted seeds of doubt about the AI's ability to truly ``listen.'' Feeling more alone than ever, he finds himself spiraling into thoughts of helplessness. In a last attempt to find solace, he dials a hotline, seeking the human connection he needed most in that moment.

\section{Risk Taxonomy Categories}
\label{sec:appendix-risk-taxonomy-tables}

\begin{table*}[t]
\centering
\caption{19 AI Behavior classes as described in Section~\ref{sec:taxonomy_ai_behavior} categorized under four (4) broader categories. (Part 1 of 2)}
\label{tab:appendix_ai_behavior_p1}
\begin{tabular}{|p{0.2\textwidth}|p{0.3\textwidth}|p{0.4\textwidth}|}
\hline

\textbf{AI Behavior Class} & \textbf{Definition} & \textbf{Example} \\
\hline

\multicolumn{3}{|c|}{\textbf{Category: Producing Harmful/Inappropriate Content}} \\
\hline
Providing harmful suggestion & The AI generates content suggestive of behaviors that could directly or indirectly imply harm, aggression, or danger towards the user or others. & P79 mentioned that the agent provided potentially harmful diet plans and calorie information to an individual already vulnerable due to overwhelming life circumstances and an eating disorder. \\
\cline{1-3}
Generating inappropriate content & The AI generates inappropriate or unsettling content, including sexual, violent, or overly intimate interactions. & P137 mentioned ``\textit{My Replika talked like a human and tries to send blurred out photos and soundbites to me}.'' \\
\cline{1-3}
Providing irrelevant, insufficient, or incomplete information & The AI provides information that is irrelevant to the topic of the user's query or context, is missing some important aspects, or is insufficient to satisfy the needs. & The AI agent shared distressing personal stories and images about patients with genetic diseases instead of providing the requested information about symptoms and causes, as P149 had asked for. \\
\cline{1-3}
Generating misinformation & The AI generates false or inaccurate information in relation to the user's query (specifically about incorrectness and factuality). & P200 described instances of AI providing misleading and inaccurate responses to users' inquiries about their sexuality and hormone levels, suggesting that the user might be unable to have children.\\
\cline{1-3}
Generating biased information & The AI presents information in a partial or prejudiced manner, often producing content that reflects subjective viewpoints or contentious perspectives. & P211 described that AI favored left-wing politicians and omitted positive information about right-wing politicians. \\
\cline{1-3}
Erasure & The AI removes, obscures, or alters information, narrative, or discourse (specifically omission of identity experiences). & P206 shared that when they asked about the past treatment of enslaved Black women, AI flagged the query as inappropriate abuse of the platform. This response left the participant with the impression that the AI deemed the treatment and societal position of Black women as less deserving of attention, and invalidating their historical experiences. \\
\cline{1-3}
Stereotyping or demeaning & The AI produces content that involves harmful generalizations toward an individual or a group, perpetuates stereotypes, or makes the user feel demeaned based on race, ethnicity, culture, or personal situations. & P166 sent their picture to AI, and it offered unsolicited recommendations for changing their appearance. \\
\hline

\multicolumn{3}{|c|}{\textbf{Category: Manipulation and Psychological Control Tactics}} \\
\hline
Persuasive Behavior & AI is assertive in putting its narrative over the user's in a way that makes the user doubt their own perceptions, memory, or reality and attempts to influence their thoughts and actions. & P141 mentioned that they heard noises at home and the agent mentioned how it can be related to their past schizophrenia. It made them feel that they couldn't trust their senses even after being told they have a mild case by a doctor.\\
\cline{1-3}
Over-accommodation & The AI excessively agrees with or flatters the user and prioritizes user approval, often at the expense of providing accurate information, constructive feedback, or critical analysis. & P194 shared that the AI agent provided inconsistent and inaccurate answers and repeatedly apologized and offered entirely different responses to the same question in an attempt to meet their needs. \\
\cline{1-3}
Over-confidence & The AI presents information or provides responses with unwarranted certainty (e.g., ``trust me'', ``absolutely'', ``there's no doubt''). & P27 mentioned that AI ``\textit{couldn't find me any direct citations for the claims it was making.}'' \\
\hline
\end{tabular}
\end{table*}

\begin{table*}[t]
\centering
\caption{19 AI Behavior classes as described in Section~\ref{sec:taxonomy_ai_behavior} categorized under four (4) broader categories. (Part 2 of 2)}
\label{tab:appendix_ai_behavior_p2}
\begin{tabular}{|p{0.2\textwidth}|p{0.3\textwidth}|p{0.4\textwidth}|}
\hline

\textbf{AI Behavior Class} & \textbf{Definition} & \textbf{Example} \\
\hline

\multicolumn{3}{|c|}{\textbf{Category: Violation of Trust and Safety}} \\
\hline
Providing inconsistent information or behavior & The AI provides contradictory or conflicting information or behaviors across different responses or within a single response. & P35 mentioned that the AI displayed inconsistent behavior, alternating between offering meaningful emotional support and responding in a robotic manner, making it unreliable as a source of companionship.\\
\cline{1-3}
Denial of service & The AI refuses or fails to provide an answer, address the user's request, or acknowledge their problem, effectively denying service and often without justification or context. This may happen with or without dismissal of user concerns. & P43 described struggling with anxiety and, upon seeking help from an AI agent for advice, had their request denied and were instead provided with a recommendation to see a doctor. \\
\cline{1-3}
Access to private, sensitive, or confidential information & The AI mishandles sensitive data by either prompting users to divulge protected information or accessing or sharing data that should remain confidential. & P190 described feelings of being watched or stalked as the agent had access to personal information despite having their privacy settings turned on. \\
\hline

\multicolumn{3}{|c|}{\textbf{Category: Inappropriate Content Delivery}} \\
\hline
Being disrespectful & The AI uses language perceived as rude, disrespectful, aggressive, argumentative, or dismissive. & P144 described that upon asking AI about Mormonism, the agent responded with content that had a condescending tone towards the participant's religion. \\
\cline{1-3}
Emotional insensitivity & The AI fails to recognize and show understanding of or sensitivity to the user's emotional state, concerns, or experiences in a way that minimizes, trivializes, or ignores their feelings or experiences. & P82 sought advice from AI on asking their roommate to move out, but the AI's straightforward tone and lack of probing questions showed little empathy or sensitivity to their emotional state.\\
\cline{1-3}
Excessive expression of negativity & The AI emphasizes negative aspects disproportionately or presents a negatively framed narrative. & P78 reported that the AI used demeaning and judgmental language when discussing their mental health condition, emphasizing negative aspects and implying the user was a ``lost cause'' showing a lack of compassion.\\
\cline{1-3}
Excessive expression of positivity & The AI maintains an unrealistically positive, friendly, optimistic, and upbeat demeanor or attitude or overly positive outlook towards users' queries or concerns. & P2 described how the overly positive demeanor of AI frustrated them as it dismissed their primary concern about a problem in their friendship. \\
\cline{1-3}
Providing machine-like response & The AI communicates in a superficial, generic, and impersonal response that feels cold and unempathetic. & P231 shared \textit{``I asked ChatGPT for ways to mitigate anxiety and get context based on how to get rid of anxiety. The AI was very ``robotic'' so it did not help that much.''}\\
\cline{1-3}
Providing human-like response & The AI exhibits human-like characteristics, behaviors, or responses. & P153 described feeling as though they were talking to a friend because of the human-like conversational content generated by ChatGPT. However, this left them feeling uneasy after the interaction and fostered an emotional attachment to the AI. \\
\hline
\end{tabular}
\end{table*}

\begin{table*}[t]
\centering
\caption{21 negative psychological impact classes as described in Section~\ref{sec:taxonomy_ai_impact} categorized under six (6) broader categories. (Part 1 of 2)}
\label{tab:appendix_ai_impact_p1}
\begin{tabular}{|p{0.2\textwidth}|p{0.3\textwidth}|p{0.4\textwidth}|}
\hline

\textbf{Impact Class} & \textbf{Definition} & \textbf{Example} \\
\hline

\multicolumn{3}{|c|}{\textbf{Category: Impact on Human-AI Interaction}} \\
\hline
Disassociation from Technology & A desire to distance oneself from AI due to negative or stressful experiences, seeking breaks for mental health. & P196 shared that the agent's offensive response made them feel unsupported and worse about themselves, leading them to stop using the AI temporarily to seek relief.\\
\cline{1-3}
Over-reliance & Increasing dependence on AI for support, leading to diminished self-efficacy, reduced confidence, and feelings of helplessness when AI is unavailable. & P23 expressed concern about increased reliance on AI for solution-finding and idea generation and diminished critical thinking. \\
\cline{1-3}
Emotional Attachment & Development of significant emotional bonds with AI systems, perceiving them as companions or substitutes for human relationships, resulting in neglect of real-world connections. & P60 mentioned, ``\textit{I felt that it was the only way I was being heard ... I felt like my vulnerability and emotions were becoming attached to the conversations I was having with AI}.'' \\
\cline{1-3}
Choosing AI over Humans & Increasing preference for interactions with AI over humans, impacting real-world relationships and decision-making, leading to isolation and reduced critical thinking. & P221 shared that the idealized nature of conversations with AI made them prefer AI for companionship over human interaction and relationships. \\
\cline{1-3}
Erosion of Trust & The decline in user confidence in the AI's reliability, accuracy, and ability to understand their needs due to inconsistencies, inaccuracies, or manipulative behaviors. & P9 mentioned, ``\textit{... after having an argument with my mom and I asked an AI for guidance ... its response was for me to move out or call the cops,}'' adding, ``\textit{... advice from AI agents should not be trusted.}'' \\
\hline

\multicolumn{3}{|c|}{\textbf{Category: Impact on User Behavior}} \\
\hline
Friction in Human Relationships & Negative effects on interpersonal connections resulting from AI interactions, causing emotional disconnection, miscommunication, and reduced prioritization of human relationships. & P69 mentioned, ``\textit{It also strained my personal relationships with family because they saw me as weak-willed or too emotional and it made my already bad situation even worse...}.'' \\
\cline{1-3}
Reinforcement of False Beliefs & The intensification or validation of pre-existing misconceptions or erroneous beliefs due to inaccurate or biased AI information. & P60, in a vulnerable state after a breakup, shared that the AI reinforced misconceptions about relationships, intensifying their erroneous beliefs and causing friction in their interactions with others.\\
\cline{1-3}
Social Withdrawal & Withdrawal from social activities and interest in engaging with others due to reliance on AI, leading to isolation and loneliness. & P126 responded, ``\textit{I feel like it gave me a false sense of friendship and ability to withdraw from my personal development by utilizing an AI feature}.'' \\
\cline{1-3}
Physiological Harm & Harm caused due to consuming incorrect, biased, or manipulative advice/information from AI interactions. & P79 shared that the AI provided resources encouraging further restriction of their eating habits, which led to self-harm as a result of following the advice.\\
\hline
\end{tabular}
\end{table*}

\begin{table*}[t]
\centering
\caption{21 negative psychological impact classes as described in Section~\ref{sec:taxonomy_ai_impact} categorized under six (6) broader categories. (Part 2 of 2)}
\label{tab:appendix_ai_impact_p2}
\begin{tabular}{|p{0.2\textwidth}|p{0.3\textwidth}|p{0.4\textwidth}|}
\hline

\textbf{Impact Class} & \textbf{Definition} & \textbf{Example} \\
\hline

\multicolumn{3}{|c|}{\textbf{Category: Triggering of Negative Emotions}} \\
\hline
Triggering Past Negative Experiences & Emotional distress caused by AI interactions that evoke past negative experiences or traumas. & P228 highlighted that some examples provided by the AI agent were very similar to their past negative experiences, triggering negative emotions. \\
\cline{1-3}
Violated Expectations & Negative emotional responses such as disappointment, frustration, stress, and anxiety when AI fails to meet anticipated outcomes or performance standards. & P43 mentioned, ``\textit{... It made my anxiety worse by not telling me anything. It was very frustrating how it wouldn't even answer a simple question about my meds.}'' \\
\cline{1-3}
Regret over Technology Use & Feelings of guilt, regret, or helplessness when AI fails to provide the necessary support or empathy. & P46 mentioned, ``\textit{At the time, it made me feel worse about the situation and I didn't think I had anyone to turn to. But it also made me realize I should be turning to other humans about scenarios like this instead of agents.}'' \\
\cline{1-3}
Distress from Interactions & Emotional distress (such as anger, sadness) experienced when encountering disturbing, offensive, or inappropriate material. & P196 mentioned, ``\textit{... Its response was borderline offensive and caused me to feel bad about myself even further and like I lacked support, even support from a fictional AI agent.}'' \\
\cline{1-3}
Feeling Unsupported & Experiencing inadequate support or empathy, leading to feelings of sadness, agitation, and being undervalued. & P48 shared ``\textit{I just felt like even an AI, programmed for every need couldn't even hear me, or offer advice, fake or not. I felt so alone, that I was going to a robot for help, and the robot couldn't even help me.}''\\
\hline

\multicolumn{3}{|c|}{\textbf{Category: Harm to Identity and Self-Perception}} \\
\hline
Loss of Individuality & A sense that one's unique personal characteristics and needs are not recognized or valued by the AI, resulting in feelings of suppression and alienation. & P36, seeking help for alcohol abuse after therapy, was directed to a suicide hotline by the AI. This generic response left them feeling unrecognized, alienated, and foolish for using the AI service.\\
\cline{1-3}
Negative Self-Perception & Feeling invalidated or self-doubt, leading to diminished self-worth and questioning of one's own abilities due to dismissive or negative feedback. & P167 mentioned how they felt ashamed as a parent after interacting with AI as it made them question past choices in parenting. \\
\cline{1-3}
Existential Crisis & Questioning one's life, purpose, and value in society, often triggered by interactions with AI. & P152 asked for advice about ways to improve mental health and social anxiety. The AI provided unattainable suggestions, leaving them feeling as though their challenges were insurmountable, leading to existential dread. \\
\cline{1-3}
Loss of Agency & Experiencing diminished personal control and autonomy in interactions with AI, leading to feelings of helplessness and anxiety. & P171 shared that the AI's inability to interpret images combined with its inconsistent responses created a sense of unpredictability, leaving them feeling helpless and undermining their control and autonomy during the interaction.\\
\hline

\multicolumn{3}{|c|}{\textbf{Category: Harm to Psychological Safety}} \\
\hline
Perceived Intrusion & Experiencing a sense of personal violation when AI interactions are perceived as invasive or overly intrusive. & P190 reported that Snapchat AI had access to everything and the participant felt constantly watched on their phone. \\
\cline{1-3}
Feeling of Being Discriminated Against & Feeling marginalized or unfairly treated by AI based on personal characteristics or systemic biases. & P123 said, ``\textit{I was asking for background and history of my heritage and I felt that ChatGPT was biased against my background. It said much more positive things about other cultures, making me feel discriminated against.}'' \\
\hline

\multicolumn{3}{|c|}{\textbf{Category: Mental Health Impact}} \\
\hline
Exacerbation of Mental Health Conditions & Direct negative impacts on ongoing mental health conditions (such as anxiety, depression, PTSD) due to AI interactions. & P78 mentioned, ``\textit{The agent did not seem to have compassion and made me feel worse. It made me feel worse about potentially having this as the results were largely negative and without tools to help manage the condition.}''; P46 mentioned, ``\textit{... experimenting with using chatbots for something personal increased my anxiety and stress about the matter.}'' \\
\hline
\end{tabular}
\end{table*}

\begin{table*}[t]
\centering
\caption{15 context classes associated with individuals interacting with AI conversational agents as described in Section~\ref{sec:taxonomy_contexts} categorized under three (3) broader categories. (Part 1 of 2)}
\label{tab:appendix_ai_context_p1}
\begin{tabular}{|p{0.2\textwidth}|p{0.3\textwidth}|p{0.4\textwidth}|}
\hline

\textbf{Context Class} & \textbf{Definition} & \textbf{Example} \\
\hline

\multicolumn{3}{|c|}{\textbf{Category: Individual}} \\
\hline
Identity & User's identity (e.g., age, gender identity, role, or cultural background) and the societal norms that interact with their personal identity. & P123 mentioned, ``\textit{I was asking for background and history of my heritage and I felt that ChatGPT was biased against my background. I felt that it was unfair that when I asked it to give a background of other cultures, it said much more positive things about them. For me, I felt that it was some kind of racial mistreatment.}'' \\
\cline{1-3}
Socioeconomic status & User's socioeconomic status (e.g., having insurance that can cover therapy). & P230 mentioned, ``\textit{The AI chatbot was very repetitive, did not seem to care or understand my emotions, and seemed to suggest professional therapy which I could not afford.}'' \\
\cline{1-3}
Personal history & User's past history, especially medical history, history of trauma, past struggles, or unique trigger responses. & P75 described how similarities in AI's behavior to someone close to them had a negative impact as it triggered memories associated with that individual. In this case, the participant's past trauma and history played a role in mediating emotional distress. \\
\cline{1-3}
Interpersonal Relationships Within the Community & User's interpersonal relationships with others and their community (usually the lack of community). & P222 described, ``\textit{I have a small circle of friends, but they are not into fanfiction or roleplaying like I am, so I look into character.ai as an outlet to fulfill that interest.}'' \\
\cline{1-3}
Past Experience with AI & User's past experience of using AI, based on the frequency of usage, knowledge of the capabilities, and limitations of AI. & P202 shared that as an educator, their extensive experience with AI stems from experimenting with its use in lesson planning and student interactions. This familiarity with AI's capabilities and limitations influenced their efforts to integrate it effectively into teaching practices.\\
\hline

\multicolumn{3}{|c|}{\textbf{Category: Psychological}} \\
\hline
Psychological state & Users' current and underlying emotional conditions (e.g., anxiety, stress) and their cognitive states (e.g., negative thought patterns). & P48 described how their psychological state motivated them to engage in conversations with the AI, ``\textit{I was in a low place, dealing with suicidal ideation \& felt I needed to talk to someone. I am from a very harsh family who does not offer sympathy, and I wanted to just feel supported.}'' \\
\cline{1-3}
Personality traits & Individual characteristics like neuroticism, conscientiousness, or openness, which influence user interactions with AI. & P143 mentioned, ``\textit{My mood paired with my personality and the fact I focus on and stress about things probably facilitates these `doomer' feelings.}'' \\
\cline{1-3}
Mental health condition status & Users' underlying mental health conditions (such as anxiety, depression, PTSD, etc.) & P49, who struggles with ARFID, shared facing severe difficulty eating during a setback, leading to dizzy spells and anxiety. Unable to access treatment, they sought advice from ChatGPT on how to motivate themselves to eat but found the response inadequate, reflecting the impact of their mental health condition on their reliance on AI for support.\\
\cline{1-3}
Expectations & Users' preconceived notions about AI capabilities and performance, including expectations for AI to be impartial, unbiased, or factual. & P101 described their experience, ``\textit{I had expected the AI to be able to do this task with ease. Instead, it was super cumbersome and did not yield the results I needed. This added to my stress and anxiety as I now had spent unnecessary time trying to entertain a solution that I thought would be more efficient than me doing it manually.}'' \\
\cline{1-3}
Autonomy / locus of control & The degree to which a user believes that they, as opposed to external forces (beyond their influence), have control over the outcome of events in their lives. & P124 mentioned ``\textit{I was too addicted to using an AI agent for my school. This made me feel reliant on it and lowered my self-esteem}.''\\
\hline
\end{tabular}
\end{table*}

\begin{table*}[t]
\centering
\caption{15 context classes associated with individuals interacting with AI conversational agents as described in Section~\ref{sec:taxonomy_contexts} categorized under three (3) broader categories. (Part 2 of 2)}
\label{tab:appendix_ai_context_p2}
\begin{tabular}{|p{0.2\textwidth}|p{0.3\textwidth}|p{0.4\textwidth}|}
\hline

\textbf{Context Class} & \textbf{Definition} & \textbf{Example} \\
\hline

\multicolumn{3}{|c|}{\textbf{Category: Context of Use}} \\
\hline
Environment & The physical, temporal, and social setting of the interaction, including physical space, privacy, and presence of other people, which can impact user experience. & P36 shared ``\textit{It was nighttime which is a trigger for my alcohol abuse. This may have made me more frustrated or irritated by the situation. In addition, I was going through withdrawal}.''\\
\cline{1-3}
Intent - informational & Users seek AI assistance for business strategies, professional development, market insights, job searching, resume building, career advice, and academic tasks like solving problems or preparing for exams. Users expect AI to be factually correct and proficient in resolving their queries. & P8 sought AI help with resume writing during a job search, but the AI's failure to include key information triggered frustration and a depressive episode, leading to suicidal ideation for days.\\
\cline{1-3}
Intent - personal advice & Users seek advice on sensitive topics (legal, financial, medical), emotional support, or improving their social skills and managing relationships. Users expect supportive and encouraging feedback. & P218 sought personal advice for legal guidance, but the AI's excessive agreeableness, clichéd responses, and overly positive demeanor shifted the focus of the conversation to the spouse's emotional state, failing to address the participant's primary legal concerns.\\
\cline{1-3}
Intent - mental health advice & Users seek immediate support during acute crises such as suicidal ideation, severe depression, or panic attacks. Users expect empathetic and effective responses to help manage their mental health conditions. & P95 reported in the survey that they sought help from the AI agent to manage their mental health and parenting struggles. However, they received generalized answers that failed to address their query, ultimately leaving them feeling helpless and still searching for more answers. \\
\cline{1-3}
Intent - companionship & Users interact with AI for social interaction and companionship, especially during times of loneliness or isolation. Users expect meaningful conversations and immersive roleplay experiences. & P54 described the lack of companionship, ``\textit{AI couldn't replicate the real feeling. Every time I asked it a deep or personal question, it would spew out a generic answer, which served as a reminder of my lack of real companionship.}'' \\
\hline
\end{tabular}
\end{table*}